\documentclass[apj]{emulateapj}
\usepackage{bm}
\newcommand{\bjdtdb}{${\rm {BJD_{TDB}}}$}
\newcommand{\bjdutc}{${\rm {BJD_{UTC}}}$}
\newcommand{\feh}{{\left[{\rm Fe}/{\rm H}\right]}}
\newcommand{\teff}{{T_{\rm eff}}}

\newcommand{\msun}{${\rm M}_\Sun$}
\newcommand{\rsun}{${\rm R}_\Sun$}

\newcommand{\mj}{${\,{\rm M}_{\rm J}}$}
\newcommand{\rj}{${\,{\rm R}_{\rm J}}$}

\newcommand{\three}{3.6$\mu$m\ }

\begin{document}

\title{Determining Empirical Stellar Masses and Radii Using Transits and Gaia Parallaxes as Illustrated by Spitzer Observations of KELT-11b}

\author{Thomas G.\ Beatty\altaffilmark{1,2}, Daniel J. Stevens\altaffilmark{3}, Karen A. Collins\altaffilmark{4,5}, Knicole D. Col\'on\altaffilmark{6}, David J. James\altaffilmark{7}, Laura Kreidberg\altaffilmark{8,9}, Joshua Pepper\altaffilmark{10}, Joseph E. Rodriguez\altaffilmark{11}, Robert J. Siverd\altaffilmark{12}, Keivan G. Stassun\altaffilmark{4,5}, John F. Kielkopf\altaffilmark{12}}

\altaffiltext{1}{Department of Astronomy \& Astrophysics, The Pennsylvania State University, 525 Davey Lab, University Park, PA 16802, USA; tbeatty@psu.edu}
\altaffiltext{2}{Center for Exoplanets and Habitable Worlds, The Pennsylvania State University, 525 Davey Lab, University Park, PA 16802, USA}
\altaffiltext{3}{Department of Astronomy, The Ohio State University, 140 W.\ 18th Ave., Columbus, OH 43210, USA}
\altaffiltext{4}{Department of Physics and Astronomy, Vanderbilt University, Nashville, TN 37235, USA}
\altaffiltext{5}{Department of Physics, Fisk University, Nashville, TN 37208, USA}
\altaffiltext{6}{NASA Goddard Space Flight Center, Greenbelt, MD 20771, USA}
\altaffiltext{7}{Department of Astronomy, University of Washington, Box 351580, Seattle, WA 98195, USA}
\altaffiltext{8}{Harvard Society of Fellows, Harvard University, 78 Mt. Auburn St., Cambridge, MA 02138, USA}
\altaffiltext{9}{Harvard-Smithsonian Center for Astrophysics, Cambridge, MA 02138, USA}
\altaffiltext{10}{Department of Physics, Lehigh University, 16 Memorial Drive East, Bethlehem, PA 18015, USA}
\altaffiltext{11}{Las Cumbres Observatory Global Telescope Network, 6740 Cortona Drive, Suite 102, Santa Barbara, CA 93117, USA}
\altaffiltext{12}{Department of Physics and Astronomy, University of Louisville, Louisville, KY 40292, USA}

\shorttitle{KELT-11b Spitzer Transit Observations}
\shortauthors{Beatty et al.}

\begin{abstract}
Using the Spitzer Space Telescope, we observed a transit at \three of KELT-11b \citep{pepper2016}. We also observed three partial planetary transits from the ground. We simultaneously fit these observations, ground-based photometry from \cite{pepper2016}, radial velocity data from \cite{pepper2016}, and an SED model utilizing catalog magnitudes and the Hipparcos parallax to the system. The only significant difference between our results and \cite{pepper2016} is that we find the orbital period is shorter by 37 seconds, $4.73610\pm0.00003$ vs. $4.73653\pm0.00006$ days, and we measure a transit center time of \bjdtdb\ $2457483.4310\pm0.0007$, which is 42 minutes earlier than predicted. Using our new photometry, we precisely measure the density of the star KELT-11 to 4\%. By combining the parallax and catalog magnitudes of the system, we are able to measure KELT-11b's radius essentially empirically. Coupled with the stellar density, this gives a parallactic mass and radius of 1.8\,\msun\ and 2.9\,\rsun, which are each approximately $1\,\sigma$ higher than the adopted model-estimated mass and radius. If we conduct the same fit using the expected parallax uncertainty from the final Gaia data release, this difference increases to $4\,\sigma$. The differences between the model and parallactic masses and radii for KELT-11 demonstrate the role that precise Gaia parallaxes, coupled with simultaneous photometric, RV, and SED fitting, can play in determining stellar and planetary parameters. With high precision photometry of transiting planets and high precision Gaia parallaxes, the parallactic mass and radius uncertainties of stars become 1\% and 3\%, respectively. TESS is expected to discover 60 to 80 systems where these measurements will be possible. These parallactic mass and radius measurements have uncertainties small enough that they may provide observational input into the stellar models themselves.
\end{abstract}


\section{Introduction}

In one way or another, almost all of our understanding of exoplanets relies upon stellar modeling. Aside from a few notable exceptions \citep[e.g.,][]{birkby2013,lockwood2014}, exoplanet observations largely rely on seeing changes in the host stars themselves, whether that is via acceleration, dimming, or movement on the sky. Since this gives us only a relative mass or a relative radius, our estimates of the planetary properties are inextricably linked to our knowledge of the host stars themselves.

Lacking observations of the stellar radius or mass, through interferometry or other means, it nevertheless has been possible to infer stellar radii via two methods. First, it is possible to estimate the radius of the primary stars in short-period eclipsing binaries by assuming tidal synchronization: then the primary's circumference is given by the orbital period of the binary divided by the spectroscopically measured $v\sin(i)$ of the primary \citep[e.g.,][]{beatty2007}. Second, one can estimate the bolometric luminosity of a planetary host star by fitting model stellar atmospheres, chosen using a spectroscopic effective temperature, to catalog magnitudes and a measured stellar parallax. This method relies upon stellar model atmospheres to determine an effective temperature estimate and to transform the apparent flux at Earth into an overall bolometric luminosity, but this transformation is relatively well understood, having been the basis of modern astronomy for the last 150 years \citep[e.g.,][]{bessel1838}. A thus derived parallactic radius estimate is therefore essentially model-independent, and can be used in determining the other system parameters.

In systems with transiting exoplanets, where the planet-to-star mass ratio is small, it is also generically possible to accurately measure the density of the central star using only the transit photometry itself \citep{seager2003}. In cases when the planet-to-star mass ratio is not negligible, one can also iteratively solve for the stellar density \citep{pal2010}. \cite{stassun2016} recently pointed out that by coupling these transit density measurements with a set of parallactic radius estimates one can make an estimate of the stellar mass. This then provides a complete set of physical parameters for the star and planet that rely on models only insofar as what is used in estimating the bolometric luminosity from the observed catalog magnitudes.

\cite{stassun2016} used the parallax measurements in the first Gaia data release \citep{lindegren2016} and the previously published transit properties of over one hundred transiting exoplanets, to estimate the masses and radii of the host stars and the planets in these systems. While the precision of these first Gaia measurements limited \cite{stassun2016}'s ability to measure masses and radii to better than 20\% to 30\%, they note that using Gaia's final data release it should be possible to make similar estimates good to 5\% to 15\%.

Until the very recent release of the first Gaia parallaxes, the faintness limit of $V<\approx9$ in the Hipparcos data set meant that this type of analysis was impossible for the majority of transiting exoplanets, and it has not been a routine practice in characterizing discoveries. Most of the large ground-based (SuperWASP, HATNet) and space-based (Kepler) surveys typically find planets around stars with magnitudes of $10<V<13$ that are too faint to have Hipparcos parallaxes. Thus, nearly all of the characterization of all of these systems has relied on estimating the parameters of the host stars using stellar models --- with a few notable exceptions \citep[e.g.,][]{bakos2010}.

The KELT transit survey, which generally targets brighter, $8<V<11$, stars, is an exception to this rule. Four of the planets discovered by KELT (KELT-2Ab \citep{beatty2012}, KELT-4b \citep{eastman2016}, KELT-7b \citep{bieryla2015}, and KELT-11b \citep{pepper2016}) are bright enough to possess Hipparcos parallaxes. In the cases of KELT-2Ab and KELT-11b, the fractional uncertainties on the measured parallaxes were low enough that modeling the stellar bolomtric flux at Earth via the spectral energy distribution (SED) of the host star provided meaningful stellar radius constraints, which were inserted as Bayesian priors on the stellar radius during the fitting process.

In the case of KELT-11b \citep[][hereafter also ``the discovery paper'']{pepper2016}, the availability of the Hipparcos parallax not only provided a constraint on the stellar radius, it also resolved a major degeneracy in fitting for the system's parameters. Specifically, the best ground-based observations using the KELT Follow-up Network had difficulty differentiating between models with a shallow, 2.5 mmag, transit, and models with a deep, 4.5 mmag transit. The different depths led to different measurements of the ingress/egress duration, hence different estimates of $a/R_*$ for the planet, and hence also to the density of the star KELT-11 as determined via the transit. Since \cite{pepper2016} used the stellar density to constrain the stellar properties, this gave considerably different absolute masses and radii (by a factor of 1.5) for both the star and planet.

KELT-11 (HD 93396) is an evolved sub-giant, with $\teff=5375$\,K and $\log(g)=3.7$\,(cgs) \citep{pepper2016}. Coincidentally, it has been spectroscopically observed since 2007 by the ``Retired A Stars'' program \citep{johnson2007} to search for radial velocity planetary companions. Due to KELT-11's relatively large stellar radius ($\sim3\,R_\odot$), a transit of KELT-11b takes almost six hours. This meant that during the follow-up observing campaign to more precisely characterize the system, a complete transit was extremely difficult to observe from the ground, and it was impossible to observe a complete transit plus anything more than a brief out-of-transit baseline using a single ground-based observatory. A multi-site network could have stitched together a complete transit with good baseline coverage, but this would have been problematic due to another effect of KELT-11's large stellar radius: KELT-11b's relatively small transit depth. This made partial transits collected in a multi-site campaign difficult to interpret, as the corrections for airmass effects could remove some or all of the transit signal. Luckily, the stellar radius implied by the Hipparcos parallax allowed the discovery paper to differentiate between these two cases, as the 4.5 mmag fit predicted a stellar radius $6\,\sigma$ larger than the parallactic radius, while the 2.5 mmag fit agreed almost perfectly. 

In an effort to observe a precise and complete transit of KELT-11b, we observed the system using the Spitzer Space Telescope. First and foremost, we wished to verify the measurements of the system parameters made by \cite{pepper2016}, and to refine them using the Spitzer photometry. Secondly, we wished to consider how accurately and precisely we can determine stellar masses and radii, and in turn the planet masses and radii, in what is essentially an entirely empirical fashion.  That process involves the combination of high-precision photometric transit observations - here with Spitzer, but in the future with TESS - that are fit simultaneously with radial velocity data, SED data, and a precise parallax.

\section{Observations and Data Reduction}

We observed a single transit of KELT-11b at \three using the IRAC camera on the Spitzer Space Telescope, using just under 15 hours of observing time. We divided our observations into two sections: the first was an initial 0.5 hour ``settling'' period to allow the initial ramp-up in the detector to occur, followed by slightly more than 14 hours of science observations. As we note below, we used a portion of the ``settling'' observations in our data analysis due to the earlier than expected transit time. To stabilize the spacecraft's pointing we used the PCRS peak-up mode with KELT-11 as the peak-up target, and we used the $32\times32$ pixel subarray read mode to decrease the image read-out time.

The science observations ran from UT 2016 April 4 1852 to UT 2016 April 5 0948.  Due to the brightness of the star KELT-11 ($K=6.122$), we took extremely short, 0.1 second, exposures to prevent the detector from saturating. This gave us a total of 409,792 images. We used the basic calibrated data (BCD) images for all of our data reduction and analysis.

Each BCD data cube contains 64 images, and records the \bjdutc\ time of the beginning of the exposure sequence (\textsc{mbjd\_obs}) and the 64 image sequence duration (\textsc{aintbeg} and \textsc{atimeend}). We calculated the \bjdutc\ mid-point of each exposure by assuming that the first exposure was taken at \textsc{mbjd\_obs} and that all 64 exposures were evenly spaced in time between \textsc{aintbeg} and \textsc{atimeend}. We converted the resulting \bjdutc\ mid-exposure times to \bjdtdb\ by adding 68.134 seconds.

We began our data reduction by background-subtracting each of our images. To do so, we masked out the central area of each image, within 15 pixels of the image center, and used this outer portion of each image to estimate that image's background. We performed three rounds of $3\,\sigma$ clipping to remove any outliers, and then fit a Gaussian to a histogram of the remaining pixels' values. We used the mean of this Gaussian fit as the mean background for each image, which we then subtracted. The average background flux was 0.015\% of KELT-11's average flux level.

\begin{figure}[t]
\vskip 0.00in
\includegraphics[width=1.1\linewidth,clip]{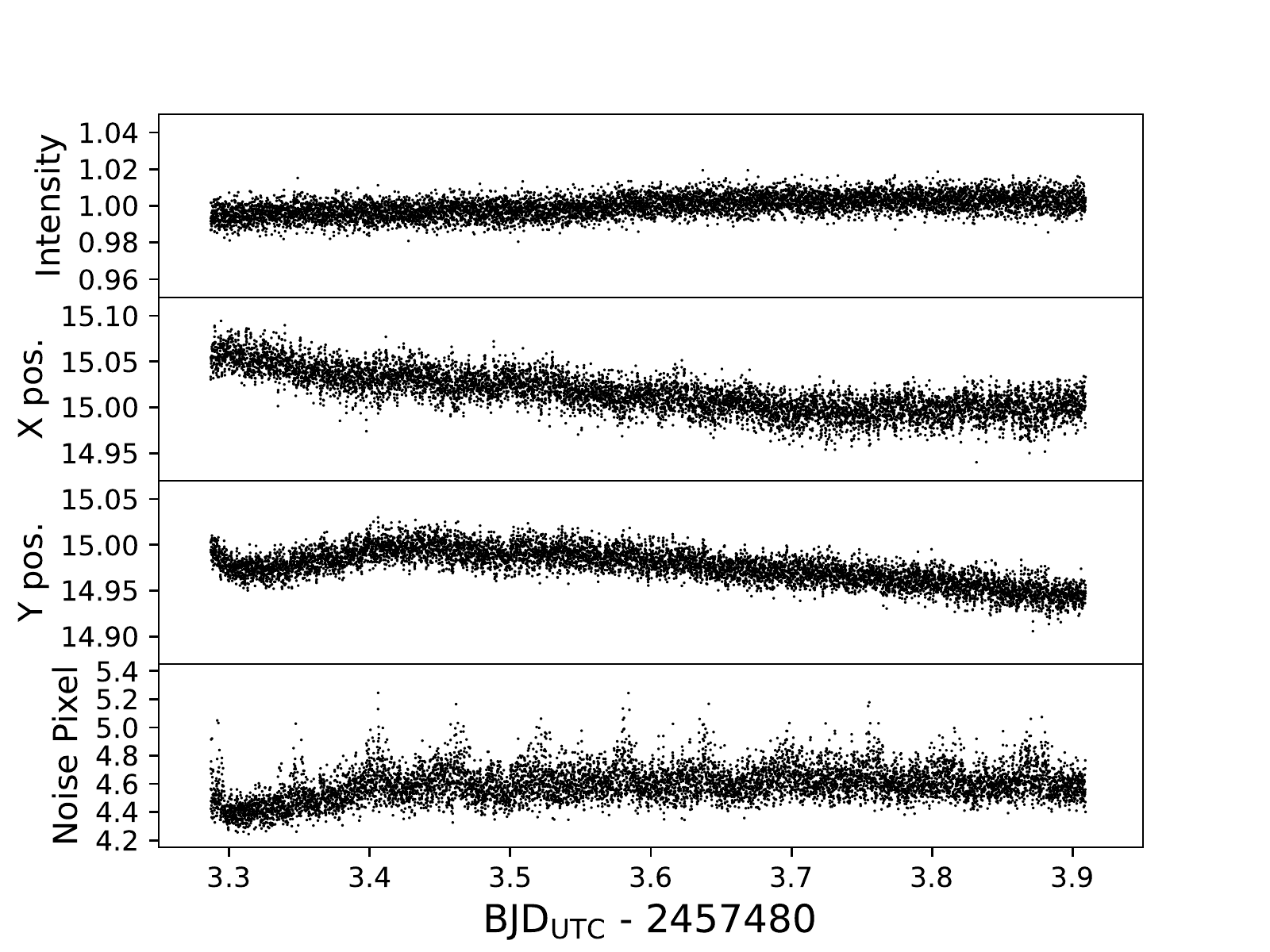}
\vskip -0.0in
\caption{Raw photometry, stellar x-position, stellar y-position, and noise pixel for our Spitzer observations. For clarity we have only plotted every 50th point.}
\label{rawphot}
\end{figure}

To help remove the well known intra-pixel effects in Spitzer/IRAC photometry, and to center our photometric extraction apertures, we measured the stellar position in each image. We did so by fitting a two-dimensional Gaussian to the stellar point-spread function (PSF) using the entire background-subtracted image. To prevent possible cosmic ray hits or other hot pixel events from generating spurious centroid shifts, we replaced any hot pixels in an image by that pixel's median flux within the image in question's 64 image data cube. To determine if a particular pixel within a particular image was ``hot'', we took each pixel's time series within a 64 image cube and flagged any images with greater than $4\,\sigma$ outliers. Note that we only conducted this hot pixel replacement for the images we used to determine the stellar centroid -- for our aperture photometry we used the uncorrected background-subtracted images.

A visual inspection of our initially extracted photometry showed that the beginning of the background-subtracted light curve still contained a noticeable ramp effect. While we would usually trim these initial images, KELT-11b's transit occurred earlier than expected in our Spitzer observations (Figure \ref{photometry}). To provide a short pre-transit baseline in the light curve, we did not trim these initial images with the ramp effect, and we also included the final 15 minutes of observations from our pre-science ``settling'' observations in our analysis. We did perform a single pass of $5\,\sigma$ clipping on our raw photometry, which removed 67 observations. This left us with 409,725 data points with which to perform our analysis (Figure \ref{rawphot}). The median photon noise error on these data was 4448\,ppm.
 
\subsection{Extraction Aperture Optimization}

To extract a light curve from our \three images, we performed simple aperture photometry on the background-subtracted images. We used the stellar centroids, as determined above, to center our extraction apertures, and we used a variable aperture size that scaled with the full-width-half-maximum (FWHM) of the stellar PSF.

\begin{figure*}[t]
\vskip 0.00in
\includegraphics[width=1.0\linewidth,clip]{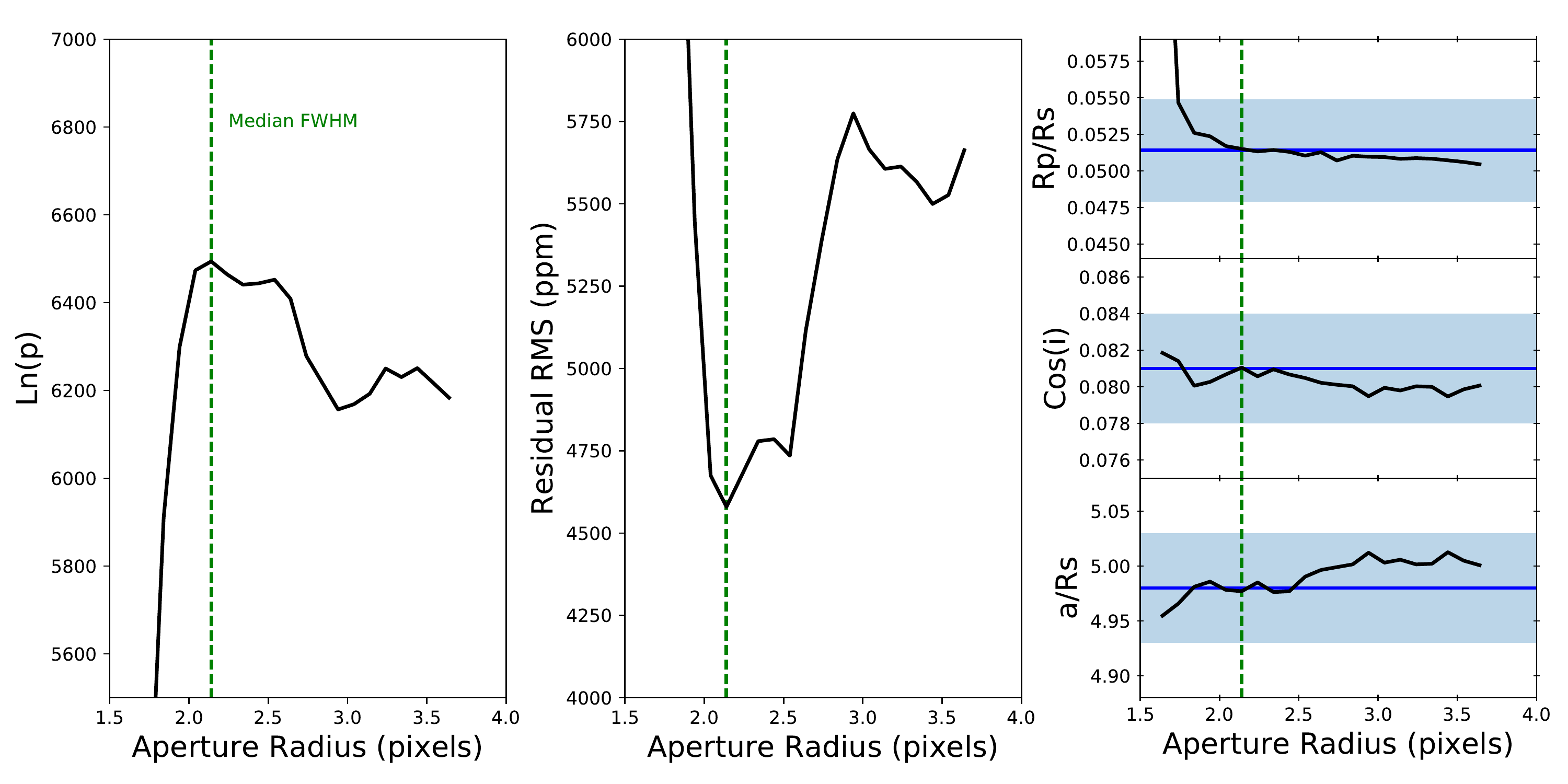}
\vskip -0.0in
\caption{The effect of differing extraction aperture radii on the initial fits to our Spitzer photometry. We selected an aperture of approximately 2.1 pixels (i.e., $C=0.0$) as the optimum, based on it having the highest $\ln(p)$ and lowest residual RMS in our initial fits. As a reminder, the actual aperture size varied as the square-root of the noise pixel parameter in each image (Equation \ref{eq:2110}). As demonstrated by the rightmost panels, the exact choice of aperture size does not significantly effect our results: the variation off all three of these parameters (black line) as a function of aperture size does not exceed the $1\,\sigma$ uncertainties we find in our final fit (blue shaded regions) -- excepting the value of $R_P/R_S$ for small apertures.}
\label{apervary}
\end{figure*}

We set our photometric extraction aperture to scale with the noise pixel parameter, which is defined in Section 2.2.2 of the \textsc{irac} Instrument Handbook v.2.1.2 as
\begin{equation}\label{eq:2110}
\beta = \frac{(\sum_i I_i)^2}{\sum_i I_i^2}.
\end{equation}
The stellar FWHM is equivalent to $\sqrt{\beta}$. To calculate the noise pixel parameter for our observations, we summed the pixels within a three pixel box of KELT-11's position -- accounting for partial pixels. KELT-11's average FWHM over our observations was 2.14 pixels, with a standard deviation of 0.03 pixels.

We set our extraction aperture to be equal to $\sqrt{\beta}+C$, where $C$ is some constant. To determine the optimal value for $C$, we extracted light curves using trial values of $C$ from -0.5 to 2.0, in steps of 0.1. We then fit each of these light curves using the initial Nelder-Mead likelihood maximization stage of our fitting process, which we describe in Section 3. below. We choose $C=0.0$ as the optimal value, as this aperture created the light curve with the highest fit likelihood, and the lowest standard deviation in the residuals to the fit (Figure \ref{apervary}). As a check to see how strongly the choice of aperture size influences our results, we examined how the fit values of $Rp/Rs$, $\cos(i)$, and $\log(a/Rs)$ varied as a function of $C$ (right panel of Figure \ref{apervary}). Aside from low values of $C$ -- which presumably are apertures not encapsulating all of the light from KELT-11 -- we found that the fit values of all three of these parameters did not vary significantly relative to our final $1\,\sigma$ uncertainties (blue shaded regions in Figure \ref{apervary}). 

\subsection{Ground-based Observations: ULMT}

\begin{figure}[t]
\vskip 0.00in
\includegraphics[width=1.1\linewidth,clip]{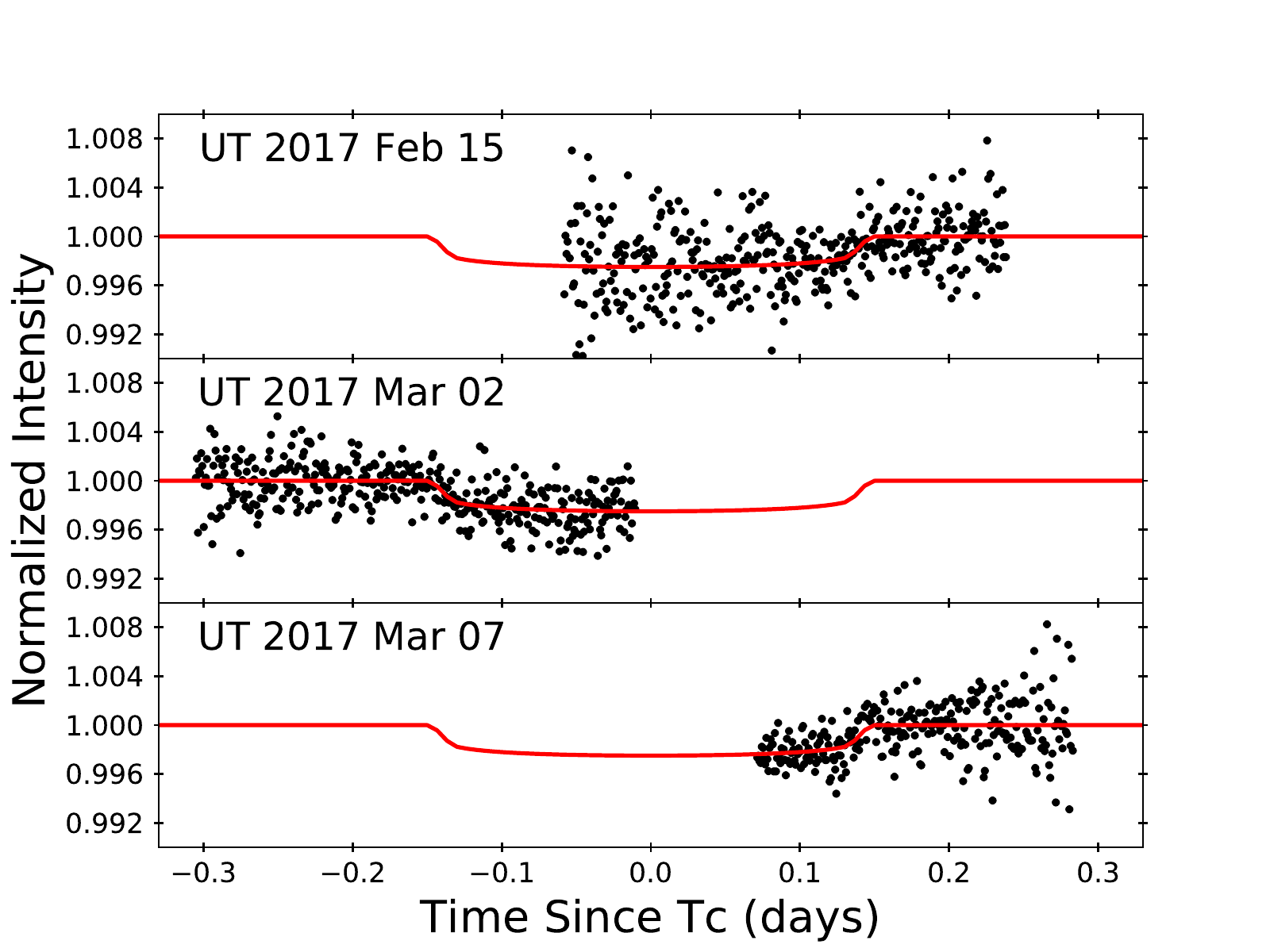}
\vskip -0.0in
\caption{In addition to the Spitzer photometry, we also observed three new partial transits of KELT-11b using the ULMT telescope (Section 2.2). Here we have plotted the detrended new ULMT light curves, along with the best fit transit model.}
\label{newground}
\end{figure}

In addition to the Spitzer observations, we also observed three partial transits of KELT-11 using the University of Louisville Manner Telescope (ULMT) at the Mt. Lemmon summit of Steward Observatory. We observed two ingresses of KELT-11b on UT 2017 February 15 and UT 2017 March 02, and an egress on UT 2017 March 07. All three observations were taken using a Sloan-$z'$ filter, and conditions were photometric throughout (Figure \ref{newground}).

The ULMT is a 0.6\,m f/8 RC Optical Systems Ritchie--Chretien, and uses  an SBIG STX-16803 4K$\times$4K CCD with 9$\,\mu$m pixels. This gives a $26'\times26'$ field of view and a plate scale of $0\farcs39$\,pixel$^{-1}$. The telescope was deliberately defocused to a FWHM of $\sim17$ pixels ($\sim6\farcs5$) for all three observations.

To calibrate the ULMT images, we used AstroImageJ \citep[AIJ,][]{collins2017} to perform bias subtraction, dark subtraction, and flat-field correction.  We also used AIJ to perform simple differential aperture photometry using a fixed aperture radius of 27 pixels to maximize flux from the star in the target and each of the three comparison star apertures while minimizing contamination from nearby stars. To determine the background level for each aperture, we used an annulus with an inner radius of 40 pixels and outer radius of 80 pixels centered on the aperture location. Outlier pixels (including pixels containing flux from stars) were removed from the background region of each aperture using an iterative $2\,\sigma$ cleaning algorithm. The average of the pixels remaining in the background region of each aperture after cleaning was then subtracted from each calibrated pixel value in the aperture. The differential photometry was then calculated as total target star flux in the target aperture divided by the sum of the total comparison star flux in the three comparison star apertures.

\section{Fitting Process}

We simultaneously fit the Spitzer photometry, the ULMT photometry, and a subset of the ground-based transit light curves used in the discovery paper, together with the radial velocity (RV) observations from the discovery paper, and a spectral energy distribution model for the star KELT-11. We discuss our selection of the discovery paper light curves at the end of Section 3.1.3. We combined all of these into one likelihood function, which we conceptually split into five parts:
\begin{eqnarray}\label{eq:3010}
\ln p_{\mathrm{tot}} &=& \ln p_{\mathrm{sptz}}+\ln p_{\mathrm{grnd}}+\ln p_{\mathrm{RV}} \\ \nonumber 
&& +\ln p_{\mathrm{SED}} + \sum\ln p_{\mathrm{priors}}.
\end{eqnarray}

We conducted the fitting process in two stages. First, we performed an initial Nelder-Mead maximization of our likelihood function to identify a starting best fit value for our model parameters. To account for the effect of unaccounted for red noise in our observations, for each of our light curves and RV data sets we scaled the median uncertainties in each individual data set to match the root mean square (RMS) standard deviation of each data set's residuals to the initial best fit. In all cases this scaling increased the median uncertainty: by a factor of 1.4 for the Spitzer observations, factors of two to ten for the ground-based photometry, and approximately a factor of two for the RV observations.

We then used the Python \textsc{emcee} package \citep{dfm2013} to perform an MCMC exploration of the parameter space surrounding the initial best fit, to determine parameter uncertainties and to verify that we had identified the global maximum likelihood value. We initialized a set of 80 walkers in parameter space around the initial best fit location, and ran each for an initial 1,000 step burn-in, before running a 10,000 step production run. At the end of the 10,000 step production run, we calculated the Gelman-Rubin statistic for each of our model parameters, and judged the MCMC chains to have converged if the Gelman-Rubin statistic was below 1.1. 

\subsection{Photometric Fitting via Non-parametric Gaussian Process Regression}

For both the Spitzer and ground-based photometry we used non-parametric Gaussian Process (GP) regression models to fit our transits of KELT-11 using the \textsc{george} Python package \citep{george2014}. Our raw Spitzer photometry displayed the usual kind of correlation for these sorts of observations: flux variation as a function of stellar centroid position, on top of a background temporal trend (Figure \ref{rawphot}).  

As mentioned in the introduction, the long duration of KELT-11b's transit meant that only one of the discovery paper light curves covered a complete transit event. None of the ULMT light curves did so. The long transit duration also meant that the one discovery light curve that was able to capture both the ingress and egress did so at high airmass ($X\approx2$). As noted in the discovery paper, this means that the discovery photometry displays strong trends with respect to airmass and time.

We chose to account for the presence of these strong correlations in our Spitzer and ground-based photometry by modeling both using a GP regression. This type of regression method models the observed data as probabilistic draws from a multivariate Gaussian distribution, rather than using a univariate Gaussian as would be done in a typical $\chi^2$ regression. This has the advantage that we may attempt to account for multi-variable correlations due to time, airmass, or stellar position directly using the multivariate Gaussian's covariance matrix, rather than projecting the airmass or stellar position correlations onto the time axis using what can be arbitrary detrending functions. \cite{rasmussen2006} gives a much more detailed mathematical description of GPs, and \cite{gibson2012} provides one of the first discussions of GPs in the context of astronomical photometric observations. GP analyses have been used before on Spitzer data by \cite{evans2015} and \cite{montet2016}, and on ground-based exoplanet observations by \cite{beatty2016}. 

To define the multivariate Gaussian underlying our GP model we used a mean function, and a covariance kernel to generate the covariance matrix. The mean function, $T(t,\phi_T)$, was the astrophysical transit model for the KELT-11 system. We used two different covariance kernels -- $\Sigma(t,\theta)$ -- for the Spitzer observations, and the ground-based observations.

\subsubsection{The Transit Model, Parameters, and Priors}

For the transit model $T(t,\phi_T)$ we used a \cite{mandel2002} transit model as implemented by the \textsc{batman} Python package \citep{kreidberg2015}. Our transit model depended upon the time, $t$, of the observations, and upon a set of physical parameters, $\phi_T$, which were the same for the Spitzer and ground-based observations. Specifically, we fit for
\begin{eqnarray}\label{eq:31110}
\phi_T &=& (T_C,\log P,\sqrt{e}\cos\omega,\sqrt{e}\sin\omega,\cos i, \\ \nonumber 
&& R_P/R_*,\log a/R_*).
\end{eqnarray}
These seven parameters are the predicted time of transit ($T_C$), the orbital period ($\log(P)$), $\sqrt{e}\cos\omega$, $\sqrt{e}\sin\omega$, the cosine of the orbital inclination ($\cos i$), the radius of the planet in stellar radii ($R_P/R_*$), the semi-major axis in units of the stellar radii ($\log[a/R_*]$). Though we median normalized our photometry, we did not explicitly fit for a baseline flux level, as this is implicitly included in the GP fitting process. Based on the lack of measured transit timing variations in the discovery paper, we also did not allow for them in our analysis. Note too that all of these parameters, except for $R_P/R_*$, are also constrained by our RV and/or SED fitting. However, we find that the photometry is the dominant constraint for most of them, except for $\sqrt{e}\cos\omega$ and $\sqrt{e}\sin\omega$.

To compare the results of the Spitzer observations to those of the discovery paper, we did not impose any priors on the transit model parameters. We did use the discovery paper values as the starting location for our initial Nelder-Mead likelihood maximization.

\subsubsection{Spitzer GP Regression Model}

As shown in Figure \ref{rawphot}, our raw Spitzer photometry displayed correlations with stellar centroid position and time that are typical for \three photometry. Following \cite{evans2015} and \cite{montet2016}, we chose to model these correlations by constructing a GP covariance kernel that was a combination of temporal and positional covariances. Specifically, we used an additive kernel
\begin{equation}\label{eq:31210}
\Sigma_{\mathrm{sptz}}(\theta_S) = \Sigma_{xyS}(\theta_S) + \Sigma_{tS}(\theta_S).
\end{equation}
We used the ``$S$'' suffix in the subscripts to denote fitting variables that related to the Spitzer data. For the positional kernel we multiplicatively combined squared exponential functions for the x- and y-position of the stellar centroid to generate the covariance between the $i$th and $j$th observations as
\begin{equation}\label{eq:31215}
\Sigma_{xyS}(\theta_S) = A_{xyS}\exp\left(-\left[\frac{x_i-x_j}{L_{xS}}\right]^2-\left[\frac{y_i-y_j}{L_{yS}}\right]^2\right),
\end{equation}
where $\theta_S=\{A_{xyS},L_{xS},L_{yS}\}$ are the hyperparameters setting the the overall amplitude ($A_{xyS}$), and the covariance length scales in x- and y-position ($L_{xS}$ and $L_{yS}$). The multiplication ensured that the GP model would only treat points as correlated if they were close to each other in both dimensions. 

For the temporal covariance kernel we used a Matern-3/2 to estimate the covariance between the $i$th and $j$th observations:
\begin{equation}\label{eq:31220}
\Sigma_{tS}(\theta_S) = A_{tS}\left(1+\frac{t_i-t_j}{L_{tS}}\sqrt{3}\right)\exp\left(-\frac{t_i-t_j}{L_{tS}}\sqrt{3}\right).
\end{equation}
The hyperparameters here, $\theta_S=\{A_{tS},L_{tS},\}$, are the overall amplitude $A_{tS}$ and the covariance length scale in time, $L_{tS}$.

The likelihood of our Spitzer GP model was then simply the likelihood of the multivariate Gaussian function defined by $\Sigma_{ij,\mathrm{sptz}}(\theta)$ and the mean function $T(t,\phi_T)$
\begin{eqnarray}\label{eq:31230}
\ln p_{\mathrm{sptz}}(\theta_G,\phi_T|f) &=& -\frac{1}{2}\bm{r}^T\Sigma_\mathrm{sptz}^{-1}\bm{r} \\ \nonumber
&& - \frac{1}{2}\log|\Sigma_\mathrm{sptz}| - \frac{N}{2}\log(2\pi),
\end{eqnarray}
where $\bm{r}=f - \mathrm{T}(t,\phi_T)$ is the vector of the residuals of our observed data ($f$) from our transit model defined in Section 3.1.1, and $N$ is the number of observations. 

The major difference between our GP model for the Spitzer observations and those used by \cite{evans2015} and \cite{montet2016} was this number, $N$, of data points used in the analysis. The matrix inversion required to compute the likelihood in Equation (\ref{eq:31230}) was computationally impossible for our full Spitzer dataset, since $\Sigma_\mathrm{sptz}$ was a $399,725\times399,725\times2$ matrix.

The problem of creating GP models for large datasets is a topic extensively covered in the machine learning literature. For our Spitzer data we used a Product of Experts (PoE) model, as described by \cite{deisenroth2015}. A PoE model evenly splits the dataset into a set of $M$ ``experts,'' such that the computations necessary to calculate the likelihood of each expert's data with respect to the model is practical. We divided our dataset into $M=400$ experts, each with approximately 1,000 data points. We chose the number of experts as a balance between analysis time and fidelity relative to a full GP analysis. We experimented with using $M=85$ experts, which was the minimum practical number given our computers' inability to invert matrices much larger than $5,000\times5,000\times2$. This experimental run took slightly under four weeks to complete (compared to about 5 hours for $M=400$), and yielded parameters and uncertainties consistent with our $M=400$ fits.

The likelihood of the PoE model was then the sum of each expert's individual likelihood, 
\begin{equation}\label{eq:31240}
\ln(p)_\mathrm{sptz} = \sum_i^M\ln(p)_{\mathrm{sptz},i}.
\end{equation}
The PoE model has the virtue of being one of the simplest treatments of large datasets in the GP framework. As shown in \cite{deisenroth2015}, a PoE model will successfully recover the same predicted mean and uncertainty on the mean as a full GP model, within the observed data range. Within the machine learning community, this has limited the appeal of PoE models, since they are typically interested in using GP techniques to predict trends outside the observed data range. For astronomical purposes, however, we are only concerned with the behavior of the mean -- the transit model -- where we have observations, and a PoE model is thus well suited to dealing with large astronomical datasets.

In our fitting we do not impose any priors on the Spitzer hyperparameters, other than requiring that the covariance length scale in time, $L_t$, be longer than the predicted durations of ingress and egress of KELT-11b's transit. This ensures that the GP regression model does not treat the transit itself as correlated noise.

\subsubsection{Ground-based GP Regression Model}

For the ground-based observations we used a GP regression model based upon covariances as a function of time and airmass. We used a squared exponential covariance kernel for both, and additively combined them as
\begin{equation}\label{eq:31310}
\Sigma_{\mathrm{grnd}}(\theta_G) = \Sigma_{XG}(\theta_G) + \Sigma_{tG}(\theta_G).
\end{equation}
We calculated the airmass covariance between the $i$th and $j$th observations as 
\begin{equation}\label{eq:31320}
\Sigma_{XG}(\theta_G) = A_{XG}\exp\left(-\left[\frac{X_i-X_j}{L_{XG}}\right]^2\right),
\end{equation}
where the hyperparameters $\theta=\{A_{XG},L_{XG},\}$ are the overall amplitude $A_{XG}$ and the covariance length scale with respect to airmass, $L_{XG}$. We used a ``$G$'' suffix to denote fitting variables relating to the ground-based light curves. We calculated the temporal covariances as
\begin{equation}\label{eq:31330}
\Sigma_{tG}(\theta_G) = A_{tG}\exp\left(-\left[\frac{t_i-t_j}{L_{tG}}\right]^2\right),
\end{equation}
where we again used an amplitude and length scale hyperparameter set, $\theta_G=\{A_{tG},L_{tG},\}$.

We used seven ground-based light curves in our fitting process. Four of these were from the discovery paper: the light curve from Westminster College Observatory (WCO) observed on UT 2015 January 01, the light curve from the Perth Exoplanet Survey Telescope (PEST) from UT 2015 March 08, the light curve from the Las Cumbres Observatory Global Telescope location in Sutherland, South Africa (CPT) on UT 2015 May 04, and the light curve observed with the Manner-Vanderbilt Ritchey-Chretien (MVRC) telescope on UT 2016 February 22.

\begin{figure*}[t]
\vskip 0.00in
\includegraphics[width=1.0\linewidth,clip]{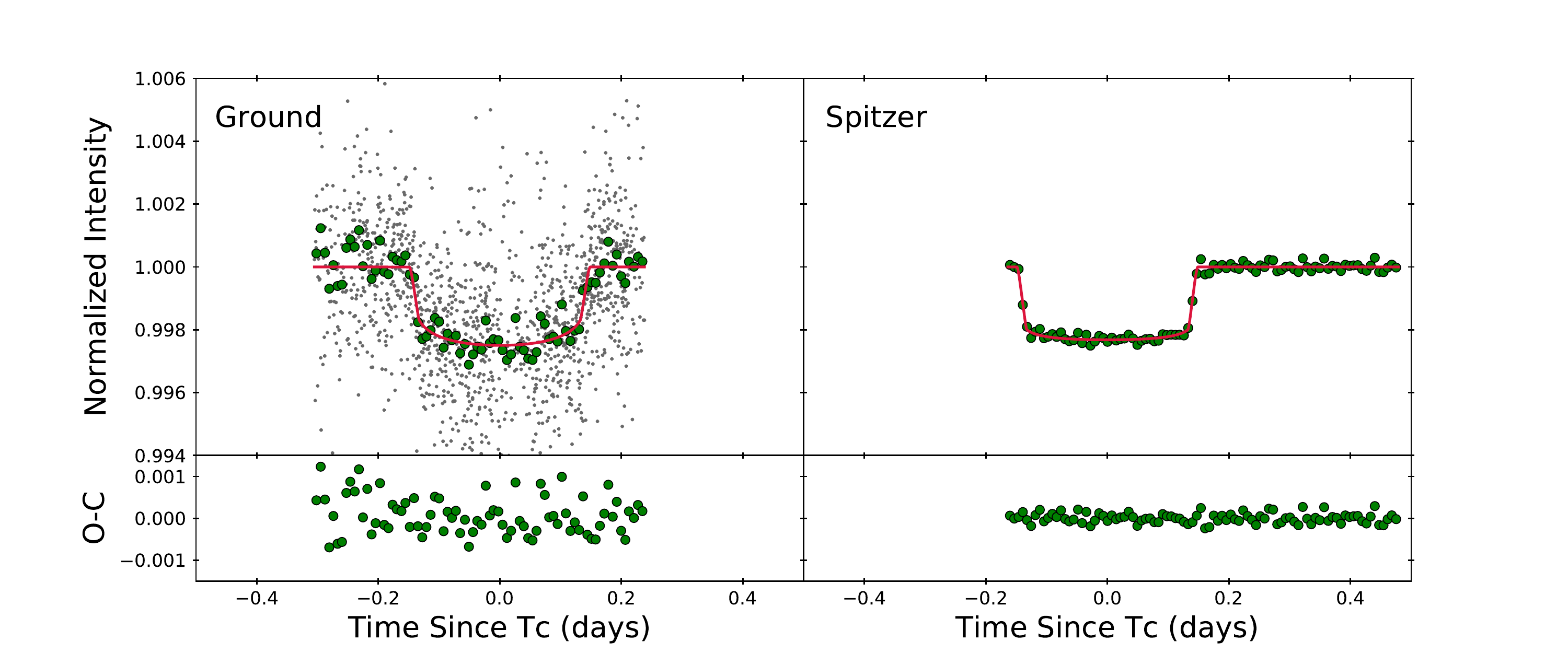}
\vskip -0.0in
\caption{The detrended ground-based and Spitzer photometry resulting from our fit using both datasets, and the Torres relations. The green points in both panels are the binned data, in 10 minute bins. The unbinned Spitzer photometry is not shown, for visual clarity.}
\label{photometry}
\end{figure*}

We selected these four discovery light curves based on the appearance of their raw, undetrended photometry. Each of them appeared to show some of the cleanest observations of the ingress and egress of KELT-11b's transit. The one exception is the MVRC light curve, which does not show an obvious transit signature, but we included because it is the only ground-based light curve in the discovery paper that saw a complete transit. 

In addition to the four discovery light curves, we also used the three partial transit observations taken using the ULMT telescope in the spring of 2017, and described in Section 2.2.

For each of the seven light curves we used individual sets of GP hyperparameters (Table \ref{tab:hyperparameters}), giving us 32 hyperparameters for the ground-based fitting. In practice, we found the bestfit hyperparameters did not vary substantially between individual light curves.  

The likelihood function we used for the ground-based GP modeling was the same as for the Spitzer data:
\begin{eqnarray}\label{eq:31340}
\ln p_{\mathrm{grnd}}(\theta_G,\phi_T|f) &=& \sum_i -\frac{1}{2}\bm{r_i}^T\Sigma_\mathrm{grnd,i}^{-1}\bm{r_i} \\ \nonumber
&& - \frac{1}{2}\log|\Sigma_\mathrm{grnd,i}| - \frac{N_i}{2}\log(2\pi),
\end{eqnarray}
where we summed each of the $i$ ground-based light curves used in our fitting process.

Similar to with the Spitzer data, we do not impose priors on the ground-based hyperparameters, other then forcing $L_{tG}$ to be longer than the ingress and egress times of the transit.

\subsection{Radial Velocity Fitting}

We used the RV measurements presented in the discovery paper for our fitting, which are one set of observations from Keck using the High Resolution Echelle Spectrometer, and another set of observations from the Automated Planet Finder (APF) using the Levy spectrograph. Since KELT-11 has been coincidentally observed as a part of the ``Retired A Stars'' program, the 17 Keck RVs span eight years, while the 15 APF RVs were observed between 2015 January 12 and 2015 November 4. The observing set-ups and the reduction process for both these datasets are described in more detail in the discovery paper. Here, we specifically use the reduced RV measurements listed in Table 16 of the discovery paper.

We fit the RV data using a single Keplerian orbit allowing for non-zero eccentricity and a background linear velocity slope. The two datasets shared the same physical parameters, but each was assigned a separate systemic velocity offset, $\gamma_{\mathrm{Keck}}$ and $\gamma_{\mathrm{APF}}$. All together, our RV model $R(t,\phi_R)$ used 10 parameters:
\begin{eqnarray}\label{eq:3210}
\phi_R &=& (M_P/M_*,\gamma_{\mathrm{Keck}},\gamma_{\mathrm{APF}},\dot{\gamma},R_*,T_C,\log P, \\ \nonumber 
&& \sqrt{e}\cos\omega,\sqrt{e}\sin\omega,\cos i,\log a/R_*).
\end{eqnarray}
In addition to the physical parameters already defined in Section 3.1.1 for our transit model, the RV model also included the planet-to-star mass ratio ($M_P/M_*$), the systemic velocities for each dataset ($\gamma_{\mathrm{Keck}}$, $\gamma_{\mathrm{APF}}$), and a linear velocity slope ($\dot{\gamma}$). We did not fit directly for the velocity semi-amplitude, $K$, of the RV orbit, but rather computed it as
\begin{equation}\label{eq:3220}
K = \frac{2\pi a_* \sin(i)}{P\sqrt{1-e^2}},
\end{equation} 
where
\begin{equation}\label{eq:3230}
a_* = \left(\frac{M_P}{M_*}\right)\left(\frac{a}{R_*}\right)R_*.
\end{equation} 
This allowed us to incorporate the observational constraints on $i$ and $a/R_*$ from the transit fits, and the constraint on $R_*$ provided by the SED fit directly into our model of the RV orbit. Similar to our transit model, we do not impose any priors on the the elements of $\phi_R$, other than using the discovery paper values as the starting point for our initial Nelder-Mead likelihood maximization. 

We calculated the likelihood of a given RV model as
\begin{equation}\label{eq:3240}
\ln p_{\mathrm{RV}} = -0.5\sum_i\left[\left(\frac{\mathrm{RV}_i-R(t_i,\phi_R)}{\sigma_{\mathrm{RV},i}}\right)^2+\ln(2\pi\sigma_{\mathrm{RV},i})\right],
\end{equation} 
over all of our $i$ RV observations.

\subsection{SED Fitting}

In addition to the photometric and RV datasets for KELT-11, we used the catalog magnitudes for the system to simultaneously fit an SED model to the stellar emission. In conjunction with the Hipparcos parallax for the system, this allowed us to determine a stellar radius for the star KELT-11 that was mostly model-independent.

\begin{figure}[t]
\vskip 0.00in
\includegraphics[width=1.1\linewidth,clip]{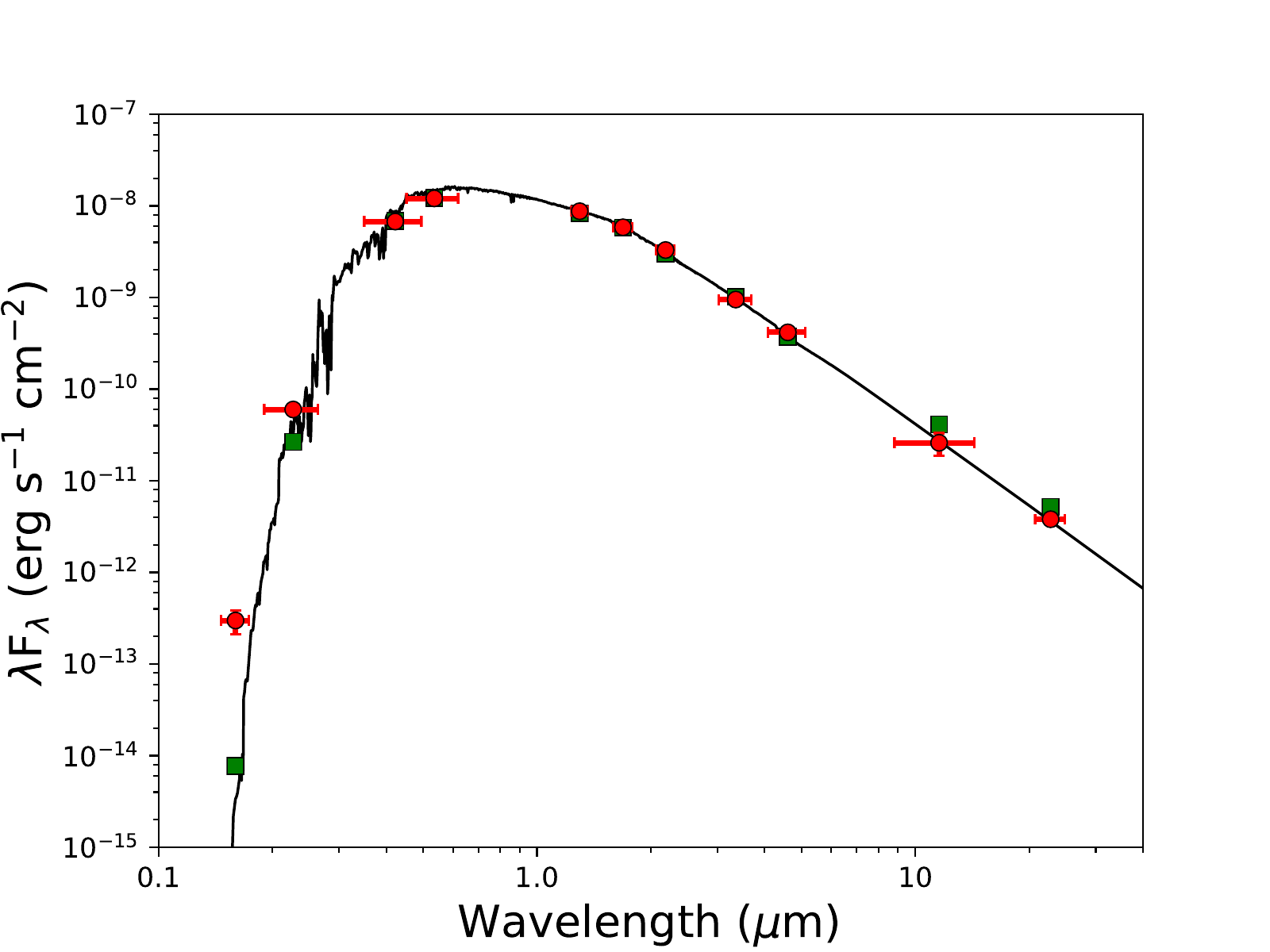}
\vskip -0.0in
\caption{The green square points show the results of our SED fit to KELT-11's catalog magnitudes (red circles). Using the Castelli-Kurucz model atmospheres \citep{ck2004}, the Hipparcos parallax, and spectroscopic constraints on KELT-11's effective temperature, surface gravity, and metallicity, we determine the stellar radius to approximately 10\%.}
\label{sedplot}
\end{figure} 

We used the GALEX, Tycho, 2MASS, and WISE magnitudes listed in Table \ref{tab:seddata} for our SED fits. To account for systematic uncertainties present in the longer wavelength WISE photometry, we increased the uncertainties on the GALEX NUV, the W3 and the W4 magnitudes from their published values of $\pm0.01$, $\pm0.015$, and $\pm0.044$, to $\pm0.1$, $\pm0.3$, and $\pm0.1$, respectively \citep{stassun2016}. 

Our SED model, $S(\phi_S)$, used five different physical parameters,
\begin{equation}\label{eq:3310}
\phi_S = (\teff,\log(g),\feh,R_*,A_V,\pi),
\end{equation}
where $A_V$ is the amount of visual extinction to KELT-11, and $\pi$ is the parallax for the system. We imposed Gaussian priors on $\teff$, $\log(g)$, and $\feh$ based on the average results of the Keck and APF spectroscopic measurements of KELT-11 listed in Table 5 of the discovery paper, with the associated $1\,\sigma$ uncertainties as the prior width. Similarly, we imposed a Gaussian prior on the parallax to KELT-11 using the Hipparocs parallax of $\pi=9.76\pm0.85$\,mas. Though KELT-11 is included in the initial Gaia source catalog, its parallax was not published as a part of the Gaia DR1 results. For the amount of visual extinction, $A_V$, to KELT-11, we did not impose a prior other than to set a maximum value of $A_V<0.1$\, magnitudes, which we based on the \cite{sfd1998} dust maps for KELT-11's position on the sky.

\begin{figure}[t]
\vskip 0.00in
\includegraphics[width=1.1\linewidth,clip]{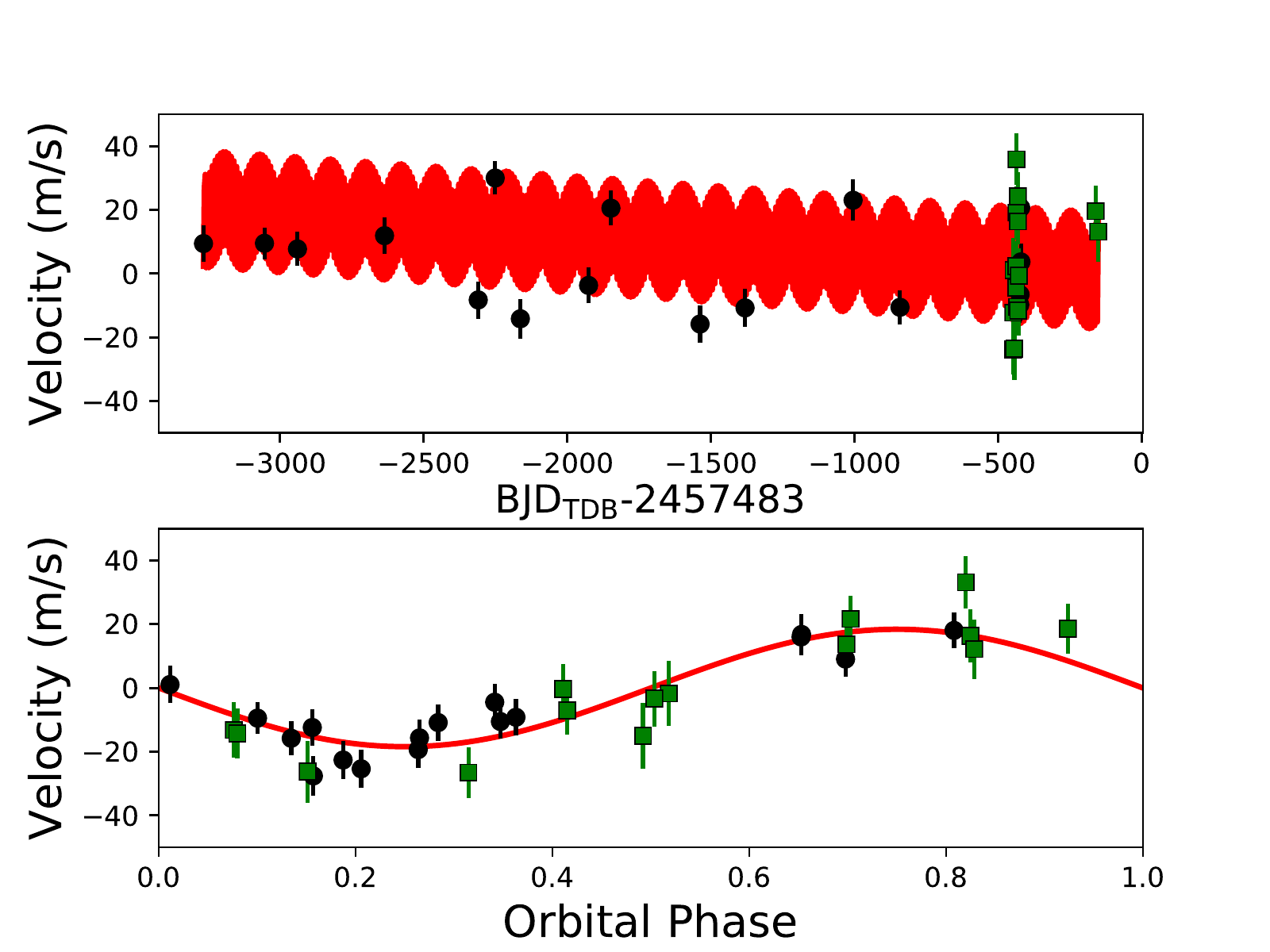}
\vskip -0.0in
\caption{The unphased and phased RV data and fit for KELT-11b, using the Spitzer and ground-based data, together with the constraints from the Torres model relations. The black round points are the Keck observations, and the green square points are from APF. The apparent amplitude modulation of the red model in the unphased plot is a plotting artifact.}
\label{rvplot}
\end{figure}

To determine the model SED for KELT-11, we used the Castelli-Kurucz model atmospheres \citep{ck2004} to compute a grid of surface luminosity magnitudes, corresponding to our catalog magnitudes, in $(\teff,\log(g),\feh)$ space. Since the Castelli-Kurucz models step by 250K in $\teff$, 0.5 dex in $\log(g)$, and 0.1 dex in $\feh$, we used a cubic spline interpolation to estimate model magnitudes in between the points provided by the Castelli-Kurucz atmospheres. We then scaled these interpolated surface magnitudes by $R_*/d$, where $d$ is the distance to the star, to determine the apparent bolometric flux of the models at Earth. We then applied a simple $R=3.1$ extinction law scaled from the value of $A_V$, to determine the extincted bolometric flux of the SED model.

Similar to the RV data, we computed the likelihood of our SED model as
\begin{equation}\label{eq:3320}
\ln(p)_{\mathrm{SED}} = -0.5\sum_i\left[\left(\frac{\mathrm{m}_i-S(\phi_S)}{\sigma_{\mathrm{m},i}}\right)^2+\ln(2\pi\sigma_{\mathrm{m},i})\right],
\end{equation} 
over all of our $i$ catalog apparent magnitudes.

\section{Results}

We fit for KELT-11's system parameters in four different ways. First, we simultaneously fit the Spitzer photometry, the RV data, and the SED model. Second, we performed a fit using both the Spitzer and the ground-based photometry, together with the RV and SED data. Third, we conducted a fit to the Spitzer, ground-based, RV, and SED data but also imposed a prior on the KELT-11's mass and radius using the Torres relations \citep{torres2010} for stellar mass and radius as a function of effective temperature, gravity, and metallicity. Fourth and finally, we did a fit to the Spitzer, ground-based, RV, and SED data, but lowered the uncertainty on the Hipparcos parallax for KELT-11 to match what is expected to be the parallax precision of the final Gaia data release. The results of all four fits are listed in Tables \ref{tab:posteriors1}, \ref{tab:posteriors2}, and \ref{tab:hyperparameters}.

We adopted the fit using the Torres relations as the true set of system parameters. As we discuss below, the measured Hipparcos parallax drives the Spitzer, and Spitzer plus Ground, fits to have values for the stellar radius higher than the expectation from stellar models. When we impose the Torres relations in our fitting process, the measured stellar density and the spectroscopic effective temperature cause the expected stellar mass and radius to be lower than in the model-free fits (e.g, Figure \ref{density}), and the change in radius causes the implied parallax to change. We consider the most likely scenario to be that the Hipparcos parallax is approximately $1\,\sigma$ lower than the true parallax value (Table \ref{tab:posteriors2}).

In terms of the photometric transit, we find that the Spitzer photometry strongly influences the fit results even when the ground-based data is included. Considering Figure \ref{photometry}, which shows both the Spitzer and ground-based photometry binned into ten minute intervals, this result is not entirely surprising given the high precision of the Spitzer data.

\section{Discussion}

\subsection{Comparison to \cite{pepper2016}}

In all of our fitting scenarios we find that KELT-11b has a shorter period, and a later transit center time, than in the discovery paper. Specifically, we find $P=4.73610\pm0.00003$ days, which is $6\,\sigma$ shorter than the discovery paper value of $P=4.73653\pm0.00006$ days. We also measure the transit center time as \bjdtdb\ $2457483.4310\pm0.0007$, which is 42 minutes ($11\,\sigma$) earlier than predicted using the discovery paper ephemeris. The rest of our measured stellar and planetary properties are consistent with the values in the discovery paper, though we measure more precise values for $\cos{i}$ and $a/R_*$.

As a test to determine the source of the orbital period discrepancy, we fit the RV orbit alone, without any photometry or SED data. This RV orbital solution gave a period of $P=4.73656$ days. We therefore consider the difference in the measured periods to not be because of any astrophysical transit timing variations, but instead  be due to the relative weight of the RV measurements in the fitting processes. In the discovery paper, the time span of the photometric observations was short enough that the RV data could ``pull'' the solution to longer periods. In our case, with transit observations two years after the discovery photometry and RV observations, the true, shorter, orbital period is required for all of the observations to line up.

\begin{figure}[b]
\vskip 0.00in
\includegraphics[width=1.1\linewidth,clip]{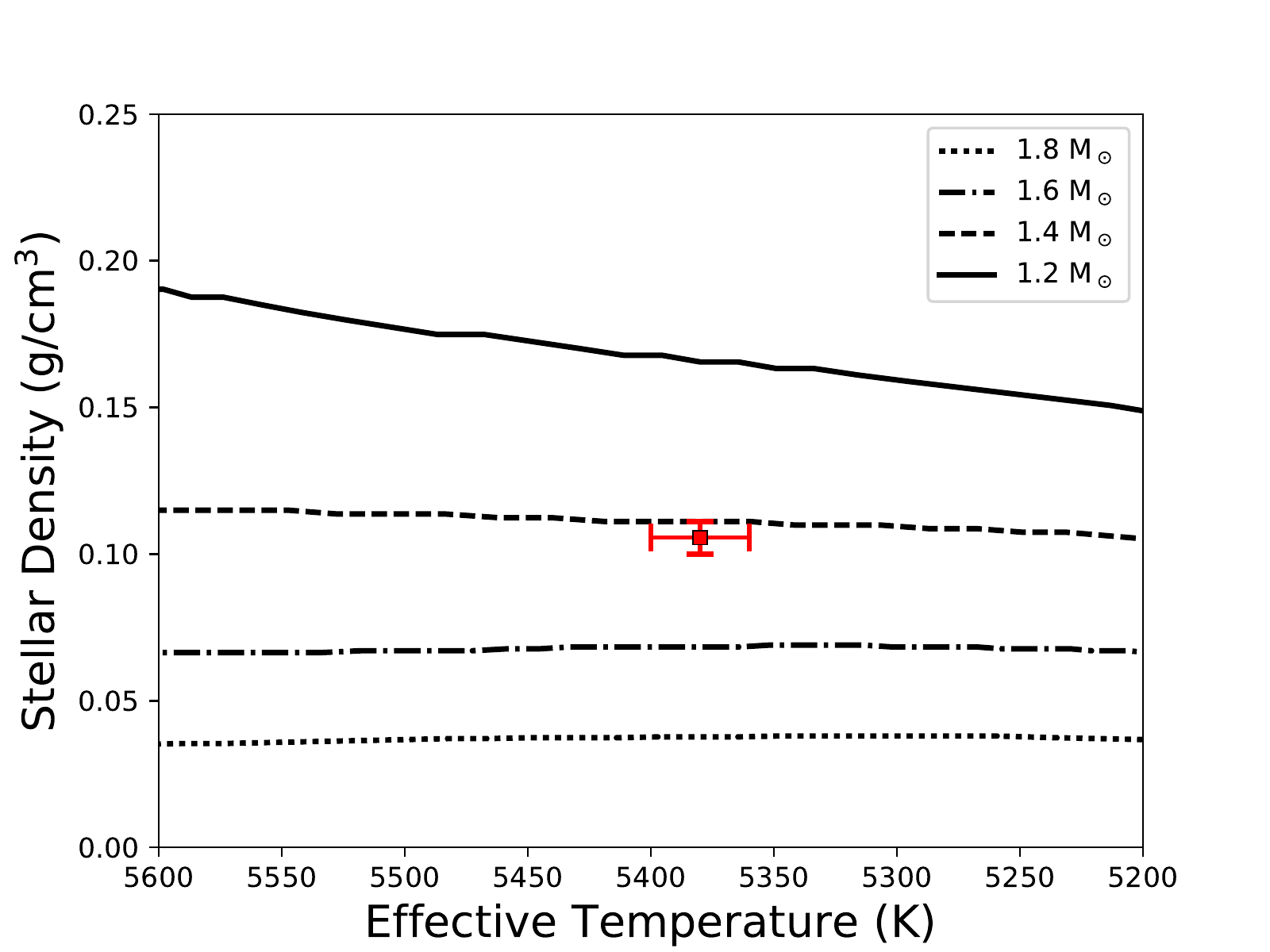}
\vskip -0.05in
\caption{The predictions of the Geneva stellar models \citep{mowlavi2012} for stellar density as a function of effective temperature and stellar mass. The stellar density is calculable directly from observables fit from the transit light curve, and our measurement of KELT-11 is the red point. Based on these models, we find the mass of KELT-11 is $1.37\pm0.03$\,\msun. This is similar to the mass found in the fit using the Torres relations, but is approximately $1\,\sigma$ lower than the fit using only parallactic constraints.}
\label{density}
\end{figure} 

\subsection{Use of Empirical Stellar Relations as Priors}

One interesting difference between the fitting cases occurs when we impose a prior on the stellar mass and radius based on the Torres relations. These relations predict the mass and radius based on the stellar effective temperature, surface gravity, and metallicity. Compared to the fit using only the Spitzer and ground-based photometry, the fit using the Torres relations finds lower mean values for the stellar radius and mass (by approximately $0.4$\,\rsun\ and $0.5$\,\msun, both about $1\,\sigma$), and a higher value for the implied stellar parallax (by 0.7\,mas, or $1\,\sigma$).

\begin{figure*}[t]
\vskip 0.00in
\includegraphics[width=1.0\linewidth,clip]{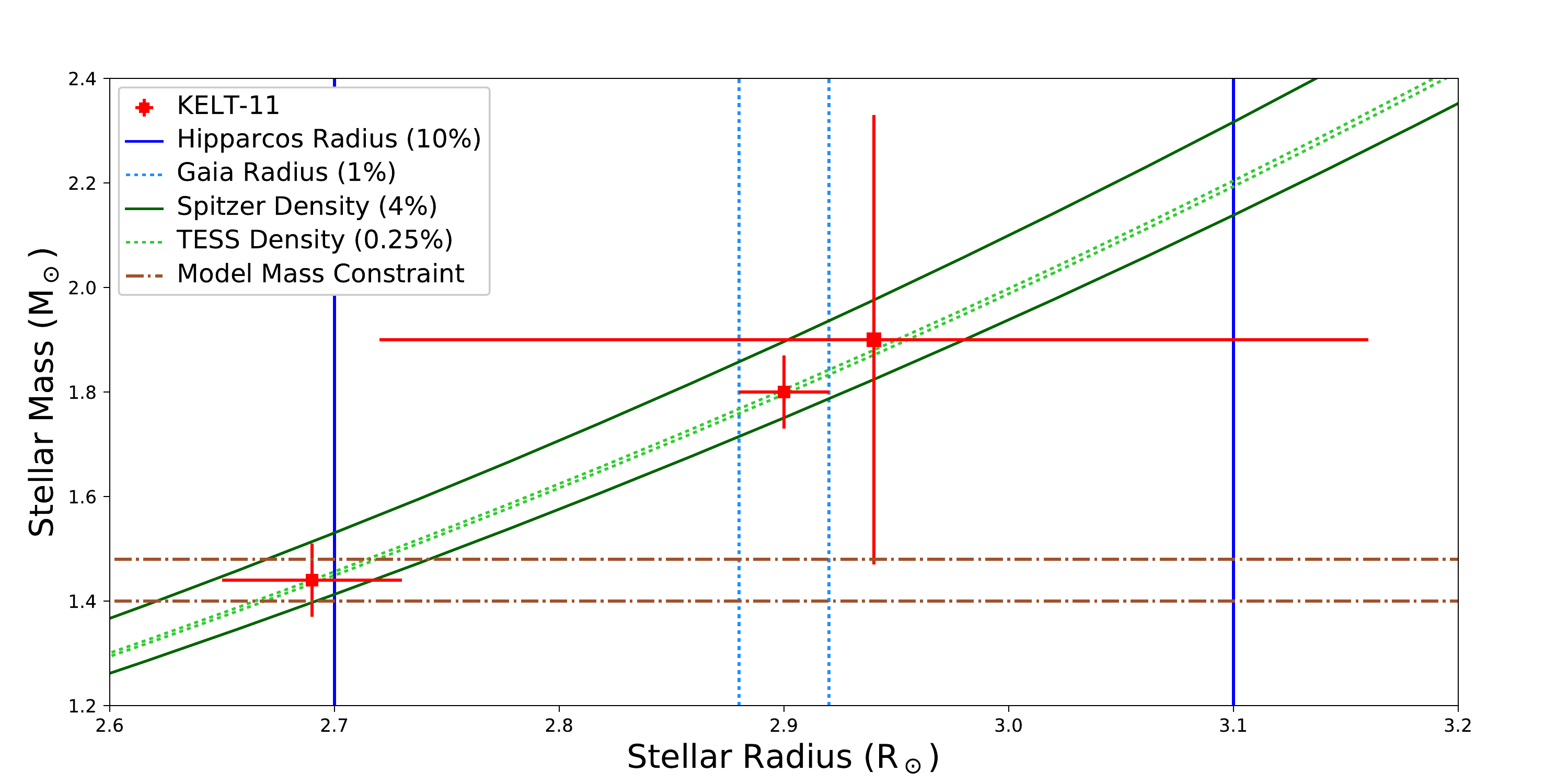}
\vskip -0.0in
\caption{Using the stellar radius determined using the Hipparcos parallax (blue solid lines) and the stellar density measured via the Spitzer light curve (solid green lines), we measure the mass and radius of KELT-11 to about 10\% (red point with large errors). The \cite{torres2010} relations impose a mass constraint based upon the spectroscopic temperature and $\log{g}$ (dash-dot brown lines), and imposing these relations upon the fit gives a resulting smaller mass and radius for KELT-11 given the measured density (leftmost red point). If we hypothetically model what will be possible with the final Gaia data release by using the Hipparcos parallax with Gaia's parallax uncertainty, we would measure a mass and radius for KELT-11 significantly different from model expectations (central red point). A hypothetical stellar density measured via TESS would make the difference even more acute. This illustrates the power of precise transit photometry and precise parallaxes to give essentially empirical stellar mass and radius measurements, with uncertainties small enough to meaningfully test stellar models. We consider the most likely stellar parameters to be the leftmost model-influenced point, under the assumptions that the \cite{torres2010} relations are correct, and that the Hipparcos parallax is $1\,\sigma$ low. A precise measurement of KELT-11's parallax via Gaia will test these assumptions.}
\label{moneyplot}
\end{figure*}

Figure \ref{density} demonstrates why this is so. Using the Geneva stellar models \citep{mowlavi2012}, we can estimate the mass of KELT-11 using the effective temperature measured via spectroscopy and the stellar density measured via the transit observables. Based on these measurements, the Geneva models predict KELT-11's mass as $1.43\pm0.03$\,\msun, which is consistent with the Torres mass determination of $1.44\pm0.07$\,\msun. Since the density of KELT-11 is determined by the transit observables, the lower mass estimate from using the stellar models leads to a lower radius estimate, and hence forces the SED fitting process to a larger implied parallax for KELT-11. 

That being said, the tension between the fits with and without the Torres relations is relatively weak: on the order of $1\,\sigma$ for all of the parameters of interest. This is almost entirely due to the 10\% fractional uncertainty on the Hipparcos parallax for KELT-11, which without the Torres relations leads directly to a 10\% fractional uncertainty on the stellar radius and a 25\% fractional uncertainty on the stellar mass. Put another way, the implied parallax necessary for the SED fits to agree with the mass and radius desired by the stellar models is not significantly different from the measured Hipparcos parallax, with $\Delta\pi=0.75\pm0.71$\,mas. Nevertheless, based on our fitting we would expect that the Gaia parallax for KELT-11 should be approximately 10.5\,mas, rather than the Hipparcos value of 9.76\,mas.

\subsection{Expected Improvements with Final Gaia Parallaxes}

To investigate how this situation would change presuming a smaller uncertainty on the measured parallax, we conducted a fit to the Spitzer and ground-based data -- without the Torres relations -- but used a hypothetical parallax measurement of $\pi=9.76\pm0.007$\,mas. This is the precision we expect for KELT-11 in the final Gaia data release \citep{lindegren2012}. As shown in the rightmost column of Table \ref{tab:posteriors2}, this leads to much smaller uncertainties in mass (by a factor of 6) and radius (by a factor of 11) for the star KELT-11. In this hypothetical scenario, the tension between the mass and radius from the stellar models, and those from the observations, becomes more acute, differing by $3.6\,\sigma$ in mass and $4.7\,\sigma$ in radius (Figure \ref{moneyplot}).

While purely notional, this illustrates the possibilities afforded by combining RV measurements, catalog magnitudes, precise photometry, and Gaia parallaxes into simultaneous fits to systems with transiting planets and eclipsing binaries. This combination will allow us to make stellar mass and radius estimations that, aside from the spectra used in the SED modeling, are largely independent of stellar models. As mentioned in the Introduction, this idea has been recently suggested both in the KELT-11 discovery paper \citep{pepper2016}, and more generally by \cite{stassun2016}. The difference between our results and those previous analyses is that we integrate the SED fitting process into our determination of the system parameters, which allows us to directly include the effects of parameter covariances in our final uncertainties, rather than assuming the covariance with the SED terms are zero \citep{pepper2016}, or using average estimated values \citep{stassun2016}. 

Nevertheless, our mass estimate for KELT-11 without using stellar models (rightmost column of Table \ref{tab:posteriors1}) is approximately a factor of two more precise than a similar model-less mass estimate from the discovery paper.\footnote{Calculable by taking the stellar radius measurements from the discovery paper together with the stellar density measurements.} This is entirely a result of our precise Spitzer photometry, which allows us to make a considerably improved measurement of $a/R_*$, and hence the stellar density, from the photometric transit. 

In the future, even more precise results should be possible by combining the final Gaia parallaxes, and the precise photometry that will be available from the TESS mission. If we presume that most of the TESS targets fall in the bright, systematics-limited, regime for Gaia parallaxes, then it should be possible to measure their stellar radii to approximately 1\% (i.e., our ``Best Gaia'' case). For a typical TESS hot Jupiter, observed at the short-cadence rate of once every minute for 27 days, if we assume the typical per-point-precision of the flux measurements are 1\,mmag \citep{sullivan2015}, then it should be possible to measure $a/R_*$ for these systems to a fractional uncertainty of approximately 0.15\% \citep{carter2008}, and the stellar density to approximately 0.25\%, or ten times better than in these observations. Coupled with the previously assumed radius uncertainties, this should allow for the masses of the hot Jupiters' host stars to be measured to 3\%, with the only model assumptions being made in the SED fitting process. For reference, an interferometric measurement of KELT-11's stellar diameter would have a precision of approximately 10\% \citep{boyajain2015}.

\subsection{The mass of KELT-11}

KELT-11 is one of the targets in the ``Retired A Stars'' search for RV planets, which uses a group of sub-giants that have putative masses near 1.5\,\msun. Though the planetary companion was not identified until the KELT observations, the statistical results using the other detections from this survey are one of the only windows we have into the planetary demographics of stars more massive than 1.5\,\msun. \cite{lloyd2011} disputed that any of the stars targeted by \cite{johnson2007}'s survey are more massive than 1.5\,\msun\ suggesting instead that the target stars are limited to between 1.0\,\msun\ to 1.2\,\msun. This has lead to a considerable amount of discussion in the literature regarding the true masses of these stars \citep{johnson2013, lloyd2013, schlaufman2013, johnson2014, ghezzi2015}. Most recently, \cite{stassun2016} made parallactic mass estimates for stars with transiting planets in the same region of the HR diagram as the ``Retired A Stars'' sample. Though these are not the actual sample stars, \cite{stassun2016} found a median mass of 1.45\,\msun, with an interquartile range of 1.22 to 1.66\,\msun. This is consistent with the stellar mass distribution for the ``Retired A Stars'' estimated by \cite{johnson2013}, which peaked near 1.3\,\msun, and higher than the mass distribution proposed by \cite{lloyd2013}, which peaked near 1.0\,\msun.

As mentioned previously, our adopted fit using the Torres relations finds that KELT-11's mass is $1.44\pm0.07$\,\msun. Based purely on the transit-derived stellar density and the star's spectroscopic effective temperature, the Geneva stellar models give an alternate mass estimate of $1.43\pm0.03$\,\msun\ (Figure \ref{density}). The probable true mass of KELT-11 is therefore approximately 1.4\,\msun, close to the 1.3\,\msun\ peak suggested by \cite{johnson2013}. 

Using the results of a single mass measurement to evaluate the ensemble properties of all of the target stars used by the ``Retired A Stars'' sample is difficult. If \cite{lloyd2013}'s mass distribution is correct, KELT-11 may just be a rare high-mass contaminant in the sample. On the other hand, \cite{johnson2013}'s mass distribution would predict that stars of KELT-11's mass would be relatively common in the ``Retired A Stars'' sample. Our measured mass for KELT-11 is therefore suggestive that \cite{johnson2013}'s higher stellar mass distribution is correct.

\section{Summary and Conclusions}

We observed a transit of KELT-11b \citep{pepper2016} at \three using the Spitzer Space Telescope, and several partial transits from the ground using the ULMT telescope. We fit our new photometry, in conjunction with some of the ground-based photometric measurements provided in \cite{pepper2016}, using a Gaussian Process regression model. We simultaneously fit these photometric observations, the RV data from \cite{pepper2016}, and an SED model utilizing catalog magnitudes and the Hipparcos parallax to the system, using a single set of stellar, orbital, and planetary parameters.

We performed four fit ``cases'' to the available observations of the KELT-11 system (Tables \ref{tab:posteriors1} and \ref{tab:posteriors2}). First, we fit just the Spitzer, RV, and SED data. Second, we fit the Spitzer observations, the ground-based observations, and the RV and SED data. Third, we used the Spitzer, ground-based, RV, and SED data, but also added modeling constraints on the KELT-11's mass and radius based on the Torres relations \citep{torres2010}. Finally, we fit the Spitzer, ground-based, RV, and SED data, but altered the uncertainty on the measured Hipparcos parallax to be match the parallax precision expected of the final Gaia data release.

Aside from the orbital ephemeris, the transit parameters we find using our Spitzer photometry are in general agreement with those given in \cite{pepper2016}. The only significant differences between our results and those of \cite{pepper2016} is that we find the orbital period to be shorter, $P=4.73610\pm0.00003$ days, vs. $P=4.73653\pm0.00006$ days, and we find the transit center occurs 42 minutes early, at \bjdtdb\ $2457483.4310\pm0.0007$.

Using the precise Spitzer photometry, we are able to measure the density of the star KELT-11 to 4\%. Using the spectroscopic effective temperature, surface gravity, and metallicity measurements of KELT-11 (Table \ref{tab:priors}), both the Torres relations (Table \ref{tab:posteriors2}) and the Geneva stellar models (Figure \ref{density}) predict that KELT-11 should be approximately 1.4\,\msun\ and 2.7\,\rsun. The Hipparcos parallax, on the other hand, gives a radius of 2.9\,\rsun, and coupled with the measured density this gives a mass of 1.8\,\msun\ (Table \ref{tab:posteriors1}). This difference is, however, not statistically significant, owing to the 10\% uncertainty on the Hipparcos parallax. Based on our fitting we would expect that the Gaia parallax for KELT-11 should be approximately 10.5\,mas, rather than the Hipparcos value of $9.76\pm0.85$\,mas.

The differences between the model and parallactic masses and radii for KELT-11 do serve to demonstrate the role that precise Gaia parallaxes, coupled with simultaneous photometric, RV, and SED fitting, can play in determining stellar parameters that are largely model-independent. To illustrate this, we also fit our KELT-11 data, but assumed a parallax error of 7\,mas (Table \ref{tab:posteriors2}). In this case, the differences between the model and hypothetical parallactic masses and radii are now $3.6\,\sigma$ and $4.7\,\sigma$, respectively. With this precision, we could begin to provide additional observational input into the stellar models themselves.

This will only improve in the future, as the short-cadence photometry available from TESS should allow us to measure the density of hot Jupiters' host stars to 0.25\%. If we combine this with the parallaxes from the final Gaia data release, then TESS should provide us with 60 to 80 stars for which we can measure the radii to 1\%, and masses to 3\%, largely independently of stellar models. This would be an improvement on current state-of-the art interferometric radius measurements, which have typical precision of 5\% to 10\% at these distances \citep{boyajain2015}. As noted in the Introduction, the measurement, interpretation, and modeling of exoplanets has relied heavily on stellar modeling. The stellar mass and radius measurements made possible by globally fitting the TESS photometry, catalog magnitudes, and Gaia parallaxes of TESS's hot Jupiters will thus allow the field of exoplanets to start paying back the debt it owes to the hard work and diligence of the stellar modeling teams.

\acknowledgments
T.G.B. was partially supported by funding from the Center for Exoplanets and Habitable Worlds. The Center for Exoplanets and Habitable Worlds is supported by the Pennsylvania State University, the Eberly College of Science, and the Pennsylvania Space Grant Consortium. Work by D.J.S was partially supported by NSF CAREER Grant AST-1056524.

This publication makes use of data products from the Two Micron All Sky Survey, which is a joint project of the University of Massachusetts and the Infrared Processing and Analysis Center/California Institute of Technology, funded by the National Aeronautics and Space Administration and the National Science Foundation.

This publication makes use of data products from the Widefield Infrared Survey Explorer, which is a joint project of the University of California, Los Angeles, and the Jet Propulsion Laboratory/California Institute of Technology, funded by the National Aeronautics and Space Administration.

This work has made use of NASA's Astrophysics Data System, the Exoplanet Orbit Database and the Exoplanet Data Explorer at exoplanets.org \citep{exoplanetsorg}, the Extrasolar Planet Encyclopedia at exoplanet.eu \citep{exoplanetseu}, the SIMBAD database operated at CDS, Strasbourg, France \citep{simbad}, and the VizieR catalog access tool, CDS, Strasbourg, France \citep{vizier}.

\begin{deluxetable*}{lcl}
\tablecaption{Values for KELT-11's System Properties from \cite{pepper2016}}
\tablehead{\colhead{~~~Parameter} & \colhead{Units} & \colhead{Value}}
\startdata
\sidehead{Stellar Properties}
~~~$\teff$\dotfill &Effective temperature (K)\dotfill & $5375\pm50$\\
~~~$\log(g_*)$\dotfill &Surface gravity (cgs)\dotfill & $3.7\pm0.1$\\
~~~$\feh$\dotfill &Metallicity\dotfill & $0.180\pm0.08$\\
~~~$\pi$\dotfill &Parallax (mas)\dotfill & $9.76\pm0.85$\\
\hline
\sidehead{Planetary Properties}
~~~$T_C$\dotfill &Time of Inferior Conjunction (\bjdtdb)\dotfill & $2457483.4605\pm0.0027$\\
~~~$P$\dotfill &Orbital period (days)\dotfill & $4.736529_{-0.000059}^{+0.000068}$\\
~~~$R_{P}/R_{*}$\dotfill &Radius of planet in stellar radii\dotfill & $0.0519\pm0.0026$\\
~~~$\cos{i}$\dotfill &Cosine of inclination\dotfill & $0.07\pm0.03$\\
~~~$a/R_{*}$\dotfill &Semi-major axis in stellar radii\dotfill & $4.93_{-0.29}^{+0.26}$\\
~~~$M_{P}/M_{*}$\dotfill &Mass ratio\dotfill & $0.000129\pm0.000012$\\
~~~$\sqrt{e}\cos{\omega}$\dotfill & \dotfill & $\equiv 0 $ \\
~~~$\sqrt{e}\sin{\omega}$\dotfill & \dotfill & $\equiv 0 $\\
~~~$\gamma_{APF}$\dotfill &m/s\dotfill & $1.9\pm2.4$\\
~~~$\gamma_{KECK}$\dotfill &m/s\dotfill & $-1.8\pm2.4$\\
~~~$\dot{\gamma}$\dotfill &RV slope (m/s/day)\dotfill & $-0.0060\pm0.0015$
\enddata
\tablecomments{We list the uncertainties on the planetary parameters for reference. In our fitting process we only impose Gaussian priors on the stellar properties, using the uncertainties above.}
\label{tab:priors}
\end{deluxetable*}

\begin{deluxetable*}{lllc}
\tablecaption{Catalog Magnitudes of the KELT-11 System}
\tablehead{\colhead{~~~Parameter} & \colhead{Description} & \colhead{Value} & \colhead{Reference}}
\startdata
\textsc{galex fuv} & \textsc{galex fuv} magnitude & $20.898\pm0.318$ & \textsc{galex mcat dr7} \\
\textsc{galex nuv} & \textsc{galex nuv} magnitude & $14.831\pm0.100$\tablenotemark{a} &\textsc{galex mcat dr7} \\
$B_T$\dotfill & Tycho $B_T$ magnitude & $9.072\pm0.018$ & \cite{tycho2} \\
$V_T$\dotfill & Tycho $V_T$ magnitude & $8.130\pm0.013$ & \cite{tycho2} \\
$J$\dotfill   & 2MASS $J$ magnitude   & $6.616\pm0.024$ & \cite{2mass} \\
$H$\dotfill   & 2MASS $H$ magnitude   & $6.251\pm0.042$ & \cite{2mass} \\
$K_S$\dotfill & 2MASS $K_S$ magnitude & $6.122\pm0.018$ & \cite{2mass} \\
\textit{W1}\dotfill & WISE magnitude\dotfill  & $6.152\pm0.100$ & \cite{wise} \\
\textit{W2}\dotfill & WISE magnitude\dotfill  & $6.068\pm0.036$ & \cite{wise} \\
\textit{W3}\dotfill & WISE magnitude\dotfill  & $6.157\pm0.300$\tablenotemark{a} & \cite{wise} \\\textit{W4}\dotfill & WISE magnitude\dotfill  & $6.088\pm0.100$\tablenotemark{a} & \cite{wise}
\enddata
\tablenotetext{a}{We increased the catalog uncertainties on the \textsc{galex nuv}, \textit{W3}, and \textit{W4} to $\pm0.1$, $\pm0.3$, and $\pm0.1$, respectively, to account for systematic uncertainties in these measurements.}
\label{tab:seddata}
\end{deluxetable*}

\begin{deluxetable*}{lcll}
\tablecaption{Fit Results}
\tablehead{\colhead{~~~Parameter} & \colhead{Units} & \colhead{Value} & \colhead{Value}}
\startdata
                                  &                                                & (Spitzer)                          & (Spitzer+Ground)                   \\
\sidehead{Stellar Properties}      
~~~$\teff$\dotfill                & Effective temperature (K)\dotfill              & $5375\pm25$                        & $5390\pm30$                        \\            
~~~$\log(g_*)$\dotfill            & Surface gravity (cgs)\dotfill                  & $3.74\pm0.1$                       & $3.7\pm0.1$                        \\            
~~~$\feh$\dotfill                 & Metallicity\dotfill                            & $0.18\pm0.08$                      & $0.17\pm0.08$                      \\            
~~~$\pi$\dotfill                  & Parallax (mas)\dotfill                         & $9.63\pm0.82$                      & $9.64\pm0.80$                      \\              
\sidehead{Derived Stellar Properties}                                                                                                                                         
~~~$\rho_*$\dotfill               & Density (cgs)\dotfill                          & $0.106\pm0.009$                    & $0.106\pm0.004$                    \\      
~~~$R_*$\dotfill                  & Radius (\rsun)\dotfill                         & $2.94\pm0.24$                      & $2.94\pm0.22$                      \\         
~~~$M_*$\dotfill                  & Mass (\msun)\dotfill                           & $1.90\pm0.50$                      & $1.90\pm0.43$                      \\        
\hline
\sidehead{Planetary Properties}
~~~$T_C$\dotfill                  & Time of inferior conjunction (\bjdtdb)\dotfill & $2457483.4311\pm0.0013$            & $2457483.4303\pm0.0008$            \\    
~~~$P$\dotfill                    & Orbital period (days)\dotfill                  & $4.7365\pm0.0002$                  & $4.73613\pm0.00004$                \\
~~~$R_{P}/R_{*}$\dotfill          & Radius of planet in stellar radii\dotfill      & $0.0503\pm0.0032$                  & $0.0502\pm0.0033$                  \\                
~~~$\cos{i}$\dotfill              & Cosine of inclination\dotfill                  & $0.083\pm0.008$                    & $0.082\pm0.005$                    \\
~~~$a/R_{*}$\dotfill              & Semi-major axis in stellar radii\dotfill       & $5.00\pm0.12$                      & $5.02\pm0.07$                      \\               
~~~$M_{P}/M_{*}$\dotfill          & Mass ratio\dotfill                             & $0.000118\pm0.000012$              & $0.000119\pm0.000013$              \\               
~~~$\sqrt{e}\cos{\omega}$\dotfill & \dotfill                                       & $-0.04\pm0.06 $                    & $-0.01\pm0.03$                     \\      
~~~$\sqrt{e}\sin{\omega}$\dotfill & \dotfill                                       & $-0.03\pm0.04 $                    & $0.02\pm0.03$                      \\                
~~~$\gamma_{APF}$\dotfill         & m/s\dotfill                                    & $-0.6\pm1.9$                       & $-1.0\pm0.9$                       \\              
~~~$\gamma_{KECK}$\dotfill        & m/s\dotfill                                    & $-5.0\pm1.4$                       & $-4.5\pm0.8$                       \\
~~~$\dot{\gamma}$\dotfill         & RV slope (m/s/day)\dotfill                     & $-0.0061\pm0.0005$                 & $-0.0060\pm0.0002$                 \\      
\sidehead{Derived Planetary Properties}    
~~~$M_{P}$\dotfill                & Mass (\mj)\dotfill                             & $0.235\pm0.067$                    & $0.237\pm0.060$                    \\         
~~~$R_{P}$\dotfill                & Radius (\rj)\dotfill                           & $1.47\pm0.15$                      & $1.47\pm0.11$                      \\     
~~~$\delta$\dotfill               & Transit depth\dotfill                          & $0.0025\pm0.0003$                  & $0.0025\pm0.0004$                  \\        
~~~$i$\dotfill                    & Inclination (degrees)\dotfill                  & $85.3\pm0.5$                       & $85.3\pm0.3$                       \\        
~~~$b$\dotfill                    & Impact parameter\dotfill                       & $0.41\pm0.04$                      & $0.412\pm0.018$                    \\             
~~~$T_{FWHM}$\dotfill             & FWHM duration (days)\dotfill                   & $0.276\pm0.01$                     & $0.2760\pm0.0040$                  \\           
~~~$\tau$\dotfill                 & Ingress/egress duration (days)\dotfill         & $0.0170\pm0.001$                   & $0.0172\pm0.0015$                  \\             
~~~$T_{14}$\dotfill               & Total duration (days)\dotfill                  & $0.293\pm0.01$                     & $0.2932\pm0.0045$                  \\              
~~~$e$\dotfill                    & Eccentricity\dotfill                           & $0.006_{-0.005}^{+0.01}$           & $0.002_{-0.0014}^{+0.005}$         \\             
~~~$\omega_*$\dotfill             & Argument of periastron (degrees)\dotfill       & $121.3_{-138.1}^{+39.6}$           & $109.0_{-171.0}^{+34.0}$           \\            
~~~$T_{P}$\dotfill                & Time of periastron (\bjdtdb)\dotfill           & $2457484.0_{-1.1}^{+0.5}$          & $2457483.7_{-2.2}^{+0.5}$          \\        
~~~$T_{S}$\dotfill                & Predicted time of eclipse (\bjdtdb)\dotfill    & $2457485.800\pm0.001$              & $2457485.7983\pm0.0008$         
\enddata
\label{tab:posteriors1}
\end{deluxetable*}

\begin{deluxetable*}{lcll}
\tablecaption{Fit Results (Continued)}
\tablehead{\colhead{~~~Parameter} & \colhead{Units} & \colhead{Value} & \colhead{Value}}
\startdata
                                  &                                                & \textbf{(Spitzer+Ground+Torres)}   & (Spitzer+Ground+Final Gaia)        \\
                                  &                                                & \textbf{(adopted)}                 & \\
\sidehead{Stellar Properties}
~~~$\teff$\dotfill                & Effective temperature (K)\dotfill              & $5375\pm25$                        & $5390\pm25$                        \\
~~~$\log(g_*)$\dotfill            & Surface gravity (cgs)\dotfill                  & $3.7\pm0.1$                        & $3.7\pm0.1$                        \\
~~~$\feh$\dotfill                 & Metallicity\dotfill                            & $0.17\pm0.07$                      & $0.17\pm0.08$                      \\            
~~~$\pi$\dotfill                  & Parallax (mas)\dotfill                         & $10.51\pm0.15$                     & $9.76\pm0.007$                     \\              
\sidehead{Derived Stellar Properties}                                                                                                                                         
~~~$\rho_*$\dotfill               & Density (cgs)\dotfill                          & $0.104\pm0.003$                    & $0.105\pm0.003$                    \\      
~~~$R_*$\dotfill                  & Radius (\rsun)\dotfill                         & $2.69\pm0.04$                      & $2.90\pm0.02$                      \\         
~~~$M_*$\dotfill                  & Mass (\msun)\dotfill                           & $1.44\pm0.07$                      & $1.80\pm0.07$                      \\        
\hline
\sidehead{Planetary Properties}
~~~$T_C$\dotfill                  & Time of inferior conjunction (\bjdtdb)\dotfill & $2457483.4310\pm0.0007$            & $2457483.4308\pm0.0007$            \\    
~~~$P$\dotfill                    & Orbital period (days)\dotfill                  & $4.73610\pm0.00003$                & $4.73612\pm0.00003$                \\
~~~$R_{P}/R_{*}$\dotfill          & Radius of planet in stellar radii\dotfill      & $0.0514_{-0.0038}^{+0.0032}$       & $0.0523\pm0.0030$                  \\                
~~~$\cos{i}$\dotfill              & Cosine of inclination\dotfill                  & $0.081\pm0.003$                    & $0.081\pm0.003$                    \\
~~~$a/R_{*}$\dotfill              & Semi-major axis in stellar radii\dotfill       & $4.98\pm0.05$                      & $4.98\pm0.06$                      \\               
~~~$M_{P}/M_{*}$\dotfill          & Mass ratio\dotfill                             & $0.000121\pm0.000009$              & $0.000118\pm0.000009$              \\               
~~~$\sqrt{e}\cos{\omega}$\dotfill & \dotfill                                       & $0.00\pm0.03$                      & $0.00\pm0.03$                      \\
~~~$\sqrt{e}\sin{\omega}$\dotfill & \dotfill                                       & $0.00\pm0.02$                      & $0.01\pm0.03$                      \\                
~~~$\gamma_{APF}$\dotfill         & m/s\dotfill                                    & $-1.0\pm0.7$                       & $-1.1\pm0.7$                       \\
~~~$\gamma_{KECK}$\dotfill        & m/s\dotfill                                    & $-4.9\pm0.6$                       & $-4.5\pm0.7$                       \\
~~~$\dot{\gamma}$\dotfill         & RV slope (m/s/day)\dotfill                     & $-0.0059\pm0.0002$                 & $-0.0059\pm0.0002$                 \\      
\sidehead{Derived Planetary Properties}
~~~$M_{P}$\dotfill                & Mass (\mj)\dotfill                             & $0.171\pm0.015$                    & $0.223\pm0.020$                    \\         
~~~$R_{P}$\dotfill                & Radius (\rj)\dotfill                           & $1.35\pm0.10$                      & $1.51\pm0.09$                      \\     
~~~$\delta$\dotfill               & Transit depth\dotfill                          & $0.00265_{-0.00038}^{+0.00035}$    & $0.0027_\pm0.0003$                 \\        
~~~$i$\dotfill                    & Inclination (degrees)\dotfill                  & $85.3\pm0.2$                       & $85.4\pm0.2$                       \\        
~~~$b$\dotfill                    & Impact parameter\dotfill                       & $0.404_{-0.018}^{+0.013}$          & $0.404\pm0.016$                    \\             
~~~$T_{FWHM}$\dotfill             & FWHM duration (days)\dotfill                   & $0.2800{-0.0037}^{+0.0033}$        & $0.279\pm0.0035$                   \\           
~~~$\tau$\dotfill                 & Ingress/egress duration (days)\dotfill         & $0.0174\pm0.0012$                  & $0.0177\pm0.0012$                  \\             
~~~$T_{14}$\dotfill               & Total duration (days)\dotfill                  & $0.2974_{-0.0039}^{+0.0034}$       & $0.2964\pm0.0037$                  \\              
~~~$e$\dotfill                    & Eccentricity\dotfill                           & $0.0007_{-0.0005}^{+0.002}$        & $0.0007_{-0.0006}^{+0.002}$        \\
~~~$\omega_*$\dotfill             & Argument of periastron (degrees)\dotfill       & $-1.0_{-74.6}^{+147.4}$            & $51.7_{-140.5}^{+92.7}$            \\
~~~$T_{P}$\dotfill                & Time of periastron (\bjdtdb)\dotfill           & $2457483.0_{-1.4}^{+1.5}$          & $2457483.6_{-2.2}^{+0.9}$          \\
~~~$T_{S}$\dotfill                & Predicted time of eclipse (\bjdtdb)\dotfill    & $2457485.7991\pm0.0007$            & $2457485.7989\pm0.0007$        
\enddata
\label{tab:posteriors2}
\end{deluxetable*}

\begin{deluxetable*}{rcccc}
\tablecaption{GP Hyperparameter Results}
\tablehead{\colhead{~~~Hyperparameter}  & \colhead{Spitzer}            & \colhead{Spitzer+Ground}     & \colhead{Spitzer+Ground+Torres} & \colhead{Spitzer+Ground+Final Gaia}}
\startdata
\sidehead{Spitzer}
~~~$A_{T}$\dotfill                      & $0.0002_{-0.00013}^{+0.005}$ & $0.0002_{-0.0002}^{+0.0021}$  & $0.0004_{-0.0004}^{+0.0022}$    & $0.0004_{-0.0004}^{+0.0022}$ \\
~~~$L_{T}$\dotfill                      & $0.71\pm0.33$                & $0.45\pm0.06$                & $0.42\pm0.08$                   & $0.43\pm0.09$                \\   
~~~$A_{xy}$\dotfill                     & $0.012\pm0.008$              & $0.003\pm0.001$              & $0.003\pm0.002$                 & $0.004\pm0.001$              \\   
~~~$L_{x}$\dotfill                      & $0.39\pm0.05$                & $0.41\pm0.05$                & $0.44\pm0.04$                   & $0.42\pm0.06$                \\
~~~$L_{y}$\dotfill                      & $0.15\pm0.09$                & $0.23\pm0.04$                & $0.21\pm0.05$                   & $0.24\pm0.04$                \\
\sidehead{Ground-based: MVRC}
~~~$A_{T}$\dotfill                      & ---                          & $0.004\pm0.004$              & $0.003_{-0.003}^{+0.009}$       & $0.004_{-0.002}^{+0.005}$    \\
~~~$L_{T}$\dotfill                      & ---                          & $0.32\pm0.05$                & $0.33\pm0.07$                   & $0.30\pm0.04$                \\
~~~$A_{X}$\dotfill                      & ---                          & ${7\times10^{-6}}_{-5\times10^{-6}}^{+0.003}$ & ${9\times10^{-6}}_{-7\times10^{-6}}^{+0.002}$ & ${9\times10^{-6}}_{-6\times10^{-6}}^{+0.001}$ \\  
~~~$L_{X}$\dotfill                      & ---                          & $0.05\pm0.02$                & $0.06\pm0.035$                  & $0.06\pm0.02$                \\
\sidehead{Ground-based: WCO}           
~~~$A_{T}$\dotfill                      & ---                          & $0.0033\pm0.0035$            & $0.005\pm0.0045$                & $0.0032\pm0.0035$            \\
~~~$L_{T}$\dotfill                      & ---                          & $0.32\pm0.06$                & $0.30\pm0.09$                   & $0.28\pm0.08$                \\
~~~$A_{X}$\dotfill                      & ---                          & $0.00013_{-0.00012}^{+0.005}$& $0.0001_{-0.00009}^{+0.004}$    & $0.00006_{-0.00005}^{+0.003}$\\
~~~$L_{X}$\dotfill                      & ---                          & $0.07\pm0.03$                & $0.07\pm0.03$                   & $0.05\pm0.02$                \\
\sidehead{Ground-based: PEST}           
~~~$A_{T}$\dotfill                      & ---                          & $0.003_{-0.002}^{+0.005}$    & $0.0047_{-0.003}^{+0.006}$      & $0.004\pm0.0035$             \\
~~~$L_{T}$\dotfill                      & ---                          & $0.34\pm0.06$                & $0.30\pm0.07$                   & $0.30\pm0.05$                \\
~~~$A_{X}$\dotfill                      & ---                          & ${2\times10^{-5}}_{-1.8\times10^{-5}}^{+0.003}$& ${2\times10^{-5}}_{-1.9\times10^{-5}}^{+0.004}$& ${3\times10^{-5}}_{-2.7\times10^{-5}}^{+0.003}$\\   
~~~$L_{X}$\dotfill                      & ---                          & $0.06\pm0.03$                & $0.05\pm0.03$                   & $0.06\pm0.025$               \\
\sidehead{Ground-based: CPT}           
~~~$A_{T}$\dotfill                      & ---                          & $0.004_{-0.003}^{+0.005}$    & $0.004\pm0.003$                 & $0.004\pm0.003$              \\
~~~$L_{T}$\dotfill                      & ---                          & $0.34\pm0.065$               & $0.31\pm0.04$                   & $0.30\pm0.03$                \\
~~~$A_{X}$\dotfill                      & ---                          & $0.004_{-0.003}^{+0.005}$    & $0.0015_{-0.0011}^{+0.0033}$    & $0.0017{-0.001}^{+0.003}$    \\
~~~$L_{X}$\dotfill                      & ---                          & $0.05\pm0.02$                & $0.06\pm0.03$                   & $0.02\pm0.02$                \\
\sidehead{Ground-based: ULMT (Feb 15)}           
~~~$A_{T}$\dotfill                      & ---                          & $0.006\pm0.005$              & $0.006\pm0.005$                 & $0.004\pm0.002$              \\
~~~$L_{T}$\dotfill                      & ---                          & $0.27\pm0.04$                & $0.27\pm0.04$                   & $0.30\pm0.03$                \\
~~~$A_{X}$\dotfill                      & ---                          & $0.0007_{-0.0006}^{+0.005}$  & $0.0005_{-0.0004}^{+0.002}$     & $0.0016_{-0.001}^{+0.003}$   \\
~~~$L_{X}$\dotfill                      & ---                          & $0.06\pm0.025$               & $0.05\pm0.01$                   & $0.06\pm0.02$                \\
\sidehead{Ground-based: ULMT (Mar 02)}           
~~~$A_{T}$\dotfill                      & ---                          & $0.003_{-0.003}^{+0.004}$    & $0.003\pm0.002$                 & $0.003_{-0.003}^{+0.002}$    \\
~~~$L_{T}$\dotfill                      & ---                          & $0.27\pm0.045$               & $0.29\pm0.035$                  & $0.33\pm0.05$                \\
~~~$A_{X}$\dotfill                      & ---                          & $0.004_{-0.003}^{+0.005}$    & $0.0014_{-0.0012}^{+0.0027}$    & $0.0013_{-0.001}^{+0.003}$   \\
~~~$L_{X}$\dotfill                      & ---                          & $0.04\pm0.02$                & $0.05\pm0.015$                  & $0.05\pm0.02$                \\
\sidehead{Ground-based: ULMT (Mar 07)}           
~~~$A_{T}$\dotfill                      & ---                          & $0.006_{-0.005}^{+0.006}$    & $0.004\pm0.003$                 & $0.003\pm0.0025$             \\
~~~$L_{T}$\dotfill                      & ---                          & $0.33\pm0.05$                & $0.29\pm0.02$                   & $0.30\pm0.03$                \\
~~~$A_{X}$\dotfill                      & ---                          & $0.004_{-0.002}^{+0.006}$    & $0.002_{-0.002}^{+0.003}$       & $0.003_{-0.002}^{+0.004}$    \\
~~~$L_{X}$\dotfill                      & ---                          & $0.06\pm0.025$               & $0.05\pm0.015$                  & $0.05\pm0.015$               
\enddata
\label{tab:hyperparameters}
\end{deluxetable*}

\end{document}